\documentclass{article}

\usepackage{arxiv}

\usepackage[numbers]{natbib}

\usepackage[utf8]{inputenc} 
\usepackage[T1]{fontenc}    
\usepackage{hyperref}       
\usepackage{url}            
\usepackage{booktabs}       
\usepackage{amsfonts}       
\usepackage{nicefrac}       
\usepackage{microtype}      
\usepackage{lipsum}
\usepackage{graphicx}
\graphicspath{ {./images/} }

\RequirePackage{etex}

\usepackage{xfrac}
\usepackage{pdfpages}
\usepackage{siunitx}
\usepackage{enumitem}
\usepackage[export]{adjustbox}
\usepackage{needspace}
\usepackage{tabularx}
\usepackage{wrapfig}
\usepackage{pdflscape}
\usepackage[referable]{threeparttablex}
\usepackage{mathptmx}
\usepackage{titlesec}
\usepackage{autonum}
\usepackage{setspace}

\titleformat{\section}
  {\normalfont\Large\bfseries}{\thesection}{1em}{}

\usepackage[labelfont = bf, justification = justified, singlelinecheck = false, figurename = Figure, tablename = Table]{caption}
\usepackage{bm}
\usepackage{soul}

\usepackage[titletoc,toc,title]{appendix}
\usepackage[position=top,singlelinecheck=off]{subfig}

\usepackage{longtable}
\usepackage{multirow}
\usepackage{booktabs} 
\usepackage{afterpage}
\usepackage{relsize}
\usepackage{cprotect}
\usepackage{fancyhdr}
\newcounter{tablenote}[table]

\usepackage{lipsum}

\usepackage{array}
\newcolumntype{P}[1]{>{\centering\arraybackslash}p{#1}}
\allowdisplaybreaks

\fancypagestyle{landscape}{
  \fancyhf{}
  \fancyfoot[C]{\thepage}
  \renewcommand{\headrulewidth}{0pt}
  
}

\title{Addressing outliers in mixed-effects logistic regression: a more robust modeling approach}

\author{
 Divan A. Burger\\
    Syneos Health, Bloemfontein, Free State, \\
    South Africa \\
    Department of Mathematical Statistics and Actuarial Science, University of the Free State, Bloemfontein, \\
    South Africa\\
  \texttt{divanaburger@gmail.com}\\
   \And
 Sean van der Merwe \\
  Department of Mathematical Statistics and Actuarial Science, University of the Free State, Bloemfontein,\\
  South Africa\\
  \And
 Emmanuel Lesaffre \\
  I-BioStat, KU Leuven, Leuven, \\
  Belgium\\
  Department of Statistics and Actuarial Science, University of Stellenbosch,\\
  South Africa\\
}

\begin{document}
\maketitle
\begin{abstract}
 This study introduces an outlier-robust model for analyzing hierarchically structured bounded count data within a Bayesian framework, utilizing a logistic regression approach implemented in \texttt{JAGS}. Our model incorporates a $t$-distributed latent variable to address overdispersion and outliers, improving robustness compared to conventional models such as the beta-binomial, binomial-logit-normal, and standard binomial models. Notably, our model targets a pseudo-median that differs from the true discrete median by less than one count; this closed-form quantity provides a robust and interpretable measure of central tendency. For comparability between all models, we additionally make predictions based on the mean proportion; however, this involves an integration step for the $t$-distributed nuisance parameter. While limited literature specifically addresses outliers in mixed models for bounded count data, this research fills that gap. The practical utility of the model is demonstrated using a longitudinal medication adherence dataset, where patient behavior often results in abrupt changes and outliers within individual trajectories. A simulation study demonstrates the binomial-logit-$t$ model's strong performance, with comparison statistics favoring it among the four evaluated models. An additional data contamination simulation confirms its robustness against outliers. Our robust approach maintains the integrity of the dataset, effectively handling outliers to provide more accurate and reliable parameter estimates.
\end{abstract}

    \section{Introduction}

    Count data are common in many fields, including epidemiology, environmental science, and social sciences. This type of data can be classified into two categories: unrestricted and bounded. Unrestricted count data, such as those modeled by the Poisson distribution, have no upper limit on the number of occurrences. The negative binomial distribution is often used when the observed variance exceeds the mean, indicating overdispersion.

    In contrast to unrestricted count data, bounded count data have a natural upper limit on the number of occurrences. This type of data is often analyzed using generalized linear models with a binomial distribution, which assumes a fixed variance-mean relationship. However, real-world bounded count data frequently exhibit overdispersion, where the observed variance exceeds the binomial model's predictions. Alternative approaches, such as the beta-binomial distribution, are employed to address this. Nonetheless, these methods may not effectively handle the random variability introduced by outliers, potentially distorting analyses and leading to incorrect conclusions.

    Recent advancements in regression models have introduced robust methods that can accommodate outliers in bounded count data. One such development is the beta-2-binomial (B2B) distribution introduced by Bayes et al. \citep{bayes2024robust}. The B2B distribution extends the traditional beta-binomial regression model to better handle overdispersion and extreme observations. It maintains the mean and variance forms of the beta-binomial model while introducing additional skewness and kurtosis, thus allowing for more robust regression modeling without increasing the complexity of interpretation. As another example, Ascari and Migliorati \citep{ascari2021new} developed the flexible beta-binomial (FBB) regression model specifically for overdispersed binomial data containing outliers and exhibiting zero inflation. Using Bayesian inference, the FBB model has been shown to outperform existing models in managing heavy tails and bimodality. Additionally, Asanya et al. \citep{asanya2023robust} introduced an outlier-robust Bayesian logistic model that utilizes an informative Student~$t$ prior distribution for regression coefficients. However, these developments have primarily been applied in simple regression contexts and have not been extensively explored within mixed model frameworks. This limitation suggests the need for specialized approaches to effectively manage the complexities introduced by outliers in mixed models for bounded count data.

    There exists literature on zero-inflated mixed-effects beta-binomial regression models designed to effectively manage overdispersion and an abundance of zeros or ones, which may be considered outliers (see, for example, the works of Wen et al. \citep{wen2024bayesian}). However, these models may not be suitable for addressing outliers that do not fall into these extreme categories: zero and one.

    We have developed a regression model based on the binomial-logit-$t$ distribution, employing a $t$-distributed latent variable to address overdispersion and outliers effectively within a mixed modeling framework. This model utilizes the logistic function to directly link the binomial probability of success to a robust $t$-distribution, parameterized by mean, scale, and degrees of freedom, thereby improving the model's robustness against outliers.

    Our framework can also be interpreted as an observation-level random effects (OLRE) logistic regression designed for overdispersed bounded count data \citep{harrison2015comparison} within mixed-effects models, incorporating both cluster-specific random effects and OLREs. OLRE models introduce random effects at the level of individual observations, unlike conventional mixed-effects models that typically include random effects at the group or cluster level. Unlike traditional OLRE models that often use Gaussian distributions to handle overdispersion, our approach incorporates a $t$-distributed latent variable as the OLRE, providing robustness to outliers in addition to overdispersion. A key advantage of our model is that the link function of covariates is directly related to the marginal pseudo-median (within one count of the true discrete median), unconditional on the OLREs, making interpretation more convenient. Because the pseudo-median is less influenced by extremes, it is often a better summary of central tendency in the presence of outliers. Additionally, the (pseudo-)median is often considered a better measure of central tendency for skewed data. However, for the mean response, one must integrate out the latent variable, adding complexity. Nonetheless, our method distinctively extends the capabilities of OLRE marginalization techniques by applying a $t$-distributed framework, thereby providing improved robustness to outliers.

    Our analysis employs Bayesian methods to implement the mixed-effects regression models, inherently marginalizing over the latent variables and thereby simplifying the estimation process. We chose to use \verb|JAGS| \citep{PLUMMER2003, DENWOOD2016} for our implementations due to its user-friendly environment, which allows for direct specification of complex Bayesian models as conceptualized, therefore obviating the need to derive asymptotic properties of estimators, a requirement in frequentist methodologies.

    Our application focuses on a longitudinal medication adherence dataset, where outliers are prevalent due to behavioral inconsistencies within individual patients. These outliers can significantly skew analysis if not adequately addressed, motivating the use of our robust modeling approach to ensure reliable estimates and insights.

    This paper is structured as follows: \autoref{sec:MOTIVATE_DATA} reintroduces the motivating dataset from a previous study, focusing on medication adherence data re-analyzed here as bounded counts data subject to outliers. \autoref{sec:BLTDIST} explains the theoretical basis of the binomial-logit-$t$ distribution. In \autoref{sec:MIXED_MODELS}, we present the mixed-effects models for bounded counts data used for application and their comparison contexts. \autoref{sec:MOD_COMP_ADE} includes the methodologies used for model evaluation, such as the Watanabe-Akaike information criterion (WAIC) for model comparison and the Kullback-Leibler (K-L) divergence for assessing model adequacy and identifying outliers. \autoref{sec:APPLICATION} applies these models to the medication adherence dataset to illustrate their practical implications. \autoref{sec:SIM_PERF_STUDY} conducts simulation studies to assess the performance and characteristics of the binomial-logit-$t$ model under various scenarios, including its robustness against outliers. Finally, \autoref{sec:DISCUSSION} discusses the findings and implications of this study, presenting conclusions.

    \section{Motivating dataset} \label{sec:MOTIVATE_DATA}

    \subsection{Dataset and methodological overview}

    This section introduces the atorvastatin dataset from the study by Vrijens et al. \citep{vrijens2006effect} which examined challenges in medication adherence data. The study assessed the impact of a pharmaceutical care program on adherence to once-daily atorvastatin, used to manage cholesterol levels and reduce cardiovascular disease risk. The study involved a control group receiving standard care and an intervention group participating in a program providing personalized counseling, medication management, and adherence support.

    Adherence was monitored using the Medication Event Monitoring System (MEMS), which records medication package openings and provides an objective measure of adherence. The study included 392 patients across both groups, with adherence data derived from electronically compiled dosing histories. Adherence was calculated monthly as the number of days the medication package was opened relative to the first pharmacy visit.

    For a more comprehensive discussion of the data from Vrijens et al. \citep{vrijens2006effect}, including specifics on the ``adherence" variable, readers are referred to Burger et al. \citep{burger2024quantile}, which provides details on the aspects of the dataset not extensively covered in this study.

    \subsection{Challenges in modeling adherence data}

    In contrast to the approach by Burger et al. \citep{burger2024quantile}, who used the Kumaraswamy distribution to model adherence as a continuous proportion within a joint quantile regression framework alongside persistence, our study views adherence as a count of adherence events. Several specific factors in the adherence data analysis drive this methodology change:
    \begin{itemize}
        \item Adherence measurements span variable observation periods, making count distributions more adaptable to such variability than continuous distributions.
        \item Considering adherence as a count better corresponds with the discrete nature of medication-taking behaviors, providing a closer match with how adherence data is gathered and understood.
        \item The Kumaraswamy distribution lacks robustness against outliers, necessitating the adoption of an approach that can effectively manage such outliers \citep{burger2023robust}.
    \end{itemize}

    \subsection{Robustness to overdispersion and outliers in adherence data}

    Adherence data often exhibit overdispersion, where the observed variability exceeds what is expected under a standard binomial distribution. This overdispersion can arise due to clustering effects, where missed medication doses on one day increase the likelihood of missed doses on subsequent days. Additionally, variations in adherence behavior across different patients and time points contribute to this phenomenon \citep{osterberg2005adherence}. Standard generalized linear mixed models based on the binomial distribution cannot adequately account for this extra variability, leading to biased parameter estimates and inference.

    In addition to overdispersion, adherence data can exhibit irregular fluctuations that are not always indicative of a patient's consistent behavior toward medication regimens \citep{gast2019medication, kim2018medication, blaschke2012adherence}. These fluctuations may stem from transient factors such as alterations in daily routines during vacations or unforeseen events like family emergencies, which disrupt usual adherence patterns. Conversely, patients who generally have poor adherence might show temporary improvements due to factors such as participation in a short-term healthcare intervention or external motivational influences. However, these improvements may not be sustained, leading to a return to lower adherence levels once the influencing circumstances cease. Such temporary deviations can introduce outliers in adherence datasets. If these outliers are not appropriately managed, they can skew the analytical results, leading to potentially misleading conclusions about adherence behaviors \citep{OSBORNE2004}. Traditional generalized linear mixed models based on the binomial distribution are not robust to these outliers, which can result in inaccurate estimates of adherence patterns and variability.

\begin{figure}[b!]
        \centering
        \includegraphics[trim=0.0cm 0.0cm 0.0cm 0.0cm, clip=true, scale=0.60]{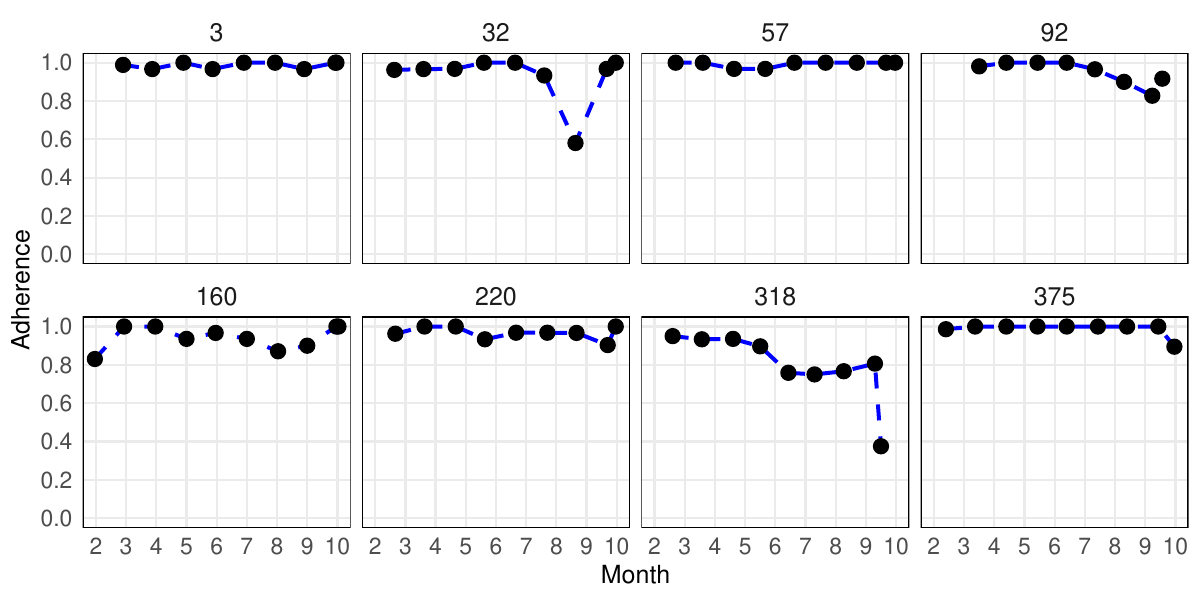}
        \caption{Adherence profiles of 8 patients completing atorvastatin treatment. Black dots indicate the proportion of days the medication was taken as prescribed each month. Blue dashed lines connect these dots, tracing adherence over time and showing strong compliance with the medication schedule.}
        \label{fig:Profile_1}
    \end{figure}

\begin{figure}[b!]
        \centering
        \includegraphics[trim=0.0cm 0.0cm 0.0cm 0.0cm, clip=true, scale=0.60]{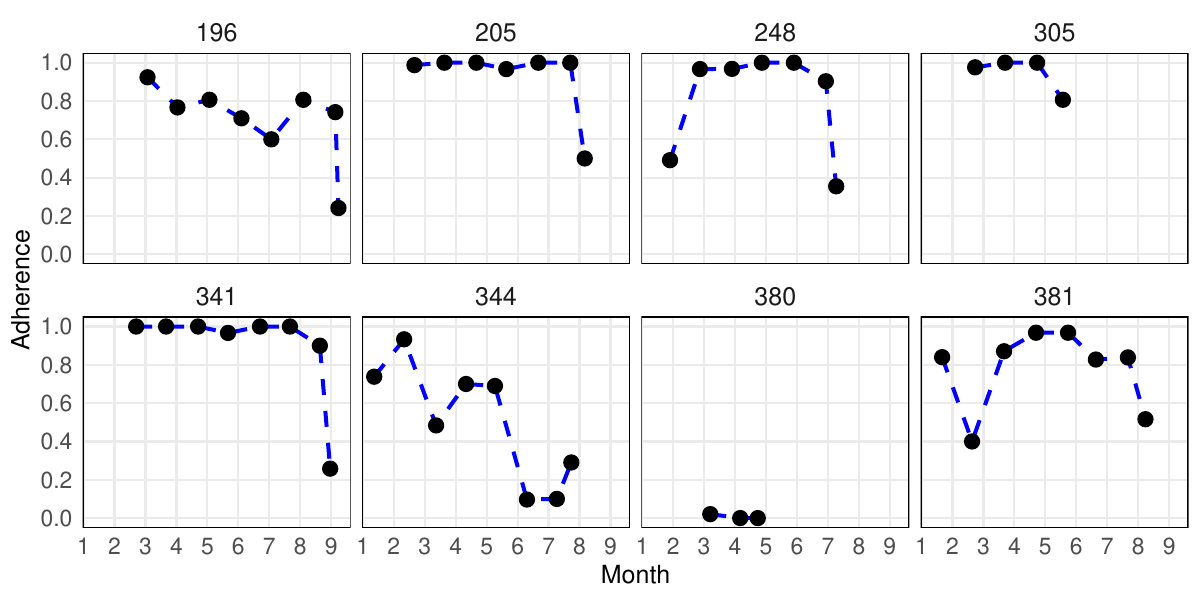}
        \caption{Adherence profiles of 8 patients who discontinued atorvastatin treatment. Black dots represent the proportion of days the medication was taken as prescribed each month. Blue dashed lines connect these dots, illustrating adherence trajectories over time, with sharp declines or significant variability before discontinuation.}
        \label{fig:Profile_2}
    \end{figure}

    \autoref{fig:Profile_1} presents a random sample of medication adherence profiles of 8 patients who completed their prescribed treatment regimen. The plots in this figure illustrate a generally high and consistent level of adherence throughout the treatment period, with occasional significant deviations that stand out but do not reflect the typical adherence behavior. \autoref{fig:Profile_2}, in contrast, presents a random sample of 8 patients who discontinued treatment prematurely. These plots reveal a notably different pattern, characterized by greater variability and several abrupt declines in adherence, culminating in complete discontinuation.

    \subsection{Motivation for a robust mixed model approach}
    
    The adherence profiles illustrated in {\color{red} Figures}~\ref{fig:Profile_1} and \ref{fig:Profile_2} show the presence of abrupt changes and outliers within individual trajectories. For our study, this motivates using a mixed model designed specifically for longitudinal data on bounded counts, which can capture the discrete nature of adherence measurements over time. Traditional continuous proportion models, such as the beta and Kumaraswamy models \citep{BAYES2017, burger2023robust}, are less suitable for this data due to the inherent bounded and discrete characteristics of adherence counts. In addition, our model needs to be able to handle outliers, which are common in datasets involving human behavior \citep{kaplan2023data}, such as drug compliance, as they can strongly influence results if not properly managed. Furthermore, overdispersion, as indicated by the variability and abrupt changes observed in the adherence data, necessitates a model that can accommodate this extra variability beyond what is expected in the standard binomial distribution. Given these considerations, we advocate for the mixed-effects binomial-logit-$t$ model, specifically designed to accommodate both overdispersion and outliers in bounded count data, providing a robust framework for analyzing medication adherence.

    \section{Binomial-logit-\textbf{\textit{t}} distribution} \label{sec:BLTDIST}

    The binomial-logit-$t$ distribution can be described as follows:
    \begin{enumerate}
        \item Let $Y_i \sim \text{Binomial}\left(m_i, p_i\right)$, where $p_i$ is the probability of success for the $i^{\text{th}}$ observation, and $m_i$ represents the number of trials.
        \item The probability $p_i$ is linked to a latent variable $\xi_i$ via the logistic function $p_i = \frac{1}{1 + e^{-\xi_i}}$.
        \item The latent variable $\xi_i$ follows a $t$-distribution parameterized by mean $\eta$, scale $\sigma^2$, and degrees of freedom $\nu$. The mean $\eta$ sets the central location, and $\sigma^2$ controls the scale, affecting dispersion. The degrees of freedom $\nu$ influence the tail behavior, impacting the distribution's robustness to outliers. For $\nu > 2$, the variance of $\xi_i$ is $\frac{\nu}{\nu-2} \sigma^2$.
    \end{enumerate}
    Given this, the probability mass function (PMF) for $Y_i = k_i$ is
    \begin{equation}
        P\left(Y_i = k_i \left| \eta, \sigma^2, \nu\right.\right) = \int_{-\infty}^{\infty} \binom{m_i}{k_i} \left(\frac{1}{1 + e^{-\xi_i}}\right)^{k_i} \left(1 - \frac{1}{1 + e^{-\xi_i}}\right)^{m_i - k_i} t\left(\xi_i \left| \eta, \sigma^2, \nu\right.\right) \, \mathrm{d}\xi_i,
    \end{equation}
    where $t\left(\xi_i \left| \eta, \sigma^2, \nu\right.\right)$ is the probability density function (PDF) of the $t$-distribution with mean $\eta$, scale parameter $\sigma^2$, and degrees of freedom $\nu$. Integrating out the nuisance parameter $\xi_i$ leads to a distribution with heavier tails. This property improves the model's robustness, effectively handling outlier values that are otherwise influential in lighter-tailed distributions.

    A convenient measure of location is obtained by passing the location parameter $\eta$ through the inverse logit and scaling by $m_i$:
    \begin{equation}\label{eq:pseudomed}
        \widetilde{M}_i = m_i \text{logit}^{-1}\left(\eta\right).
    \end{equation}
    Because $\eta$ is the median of $\xi_i$ and the logistic function is strictly increasing, $\text{logit}^{-1}\left(\eta\right)$ is the median of the latent probability $p_i$. Hence, Equation~(\ref{eq:pseudomed}) acts as a pseudo-median for the mixture-distributed count $Y_i$. It is not, in general, the exact discrete median, but the appendix shows:
    \begin{equation}
        k_{i-} \le \widetilde{M}_i < k_{i-} + 1,
    \end{equation}
    where $k_{i-}$ is the smallest integer with $P\left(Y_i \le k_{i-}\right) \ge 0.5$. We, therefore, refer to $\widetilde{M}_i$ as a ``pseudo-median" of $Y_i$, because it preserves the latent median on the probability scale while forgoing the integer constraint on $Y_i$. We adopt $\widetilde{M}_i$ throughout for interpreting the fixed-effects location, since it remains both closed-form and numerically faithful.

    More generally, any latent distribution whose median equals $\eta$ will yield the same pseudo-median property; the $t$ family is adopted here solely for its adjustable heavy tails.

    The expected value $\mathbb{E}\left[Y_i\right]$ can be computed in two steps:

    First, compute the expected value of $Y_i$ given the latent variable $\xi_i$:
    \begin{equation}
        \mathbb{E}\left[Y_i \left| \xi_i\right.\right] = m_i p_i = m_i \frac{1}{1 + e^{-\xi_i}}.
    \end{equation}
    Then, integrate this conditional expectation over the $t$-distribution of $\xi_i$:
    \begin{equation} \label{eq:MEAN_BLT}
        \mathbb{E}\left[Y_i\right] = \int_{-\infty}^{\infty} \mathbb{E}\left[Y_i \left| \xi_i \right.\right] t\left(\xi_i \left|\eta, \sigma^2, \nu \right.\right) \, \mathrm{d}\xi_i.
    \end{equation}
    Calculating the expected value $\mathbb{E}\left[Y_i\right]$ requires using numerical integration or simulation-based approaches to estimate values, as there is no closed-form solution due to integration over the latent variable $\xi_i$.

    As the variance parameter $\sigma^2$ approaches zero and the degrees of freedom $\nu$ tend to infinity in the binomial-logit-$t$ distribution, the distribution of the nuisance parameter $\xi_i$ converges to a degenerate distribution centered at $\eta$. This convergence effectively eliminates the variability introduced by $\xi_i$, reducing the distribution to the conventional binomial model. Under these circumstances, the probability of success is directly determined by the logistic function applied to $\eta$, leading to a simplification of the expected value calculation, such that $\mathbb{E}\left[Y_i\right] = m_i \frac{1}{1 + e^{-\eta}}$.

\begin{figure}[b!]
        \centering
        \includegraphics[trim=0.0cm 0.0cm 0.0cm 0.0cm, clip=true, scale=0.70]{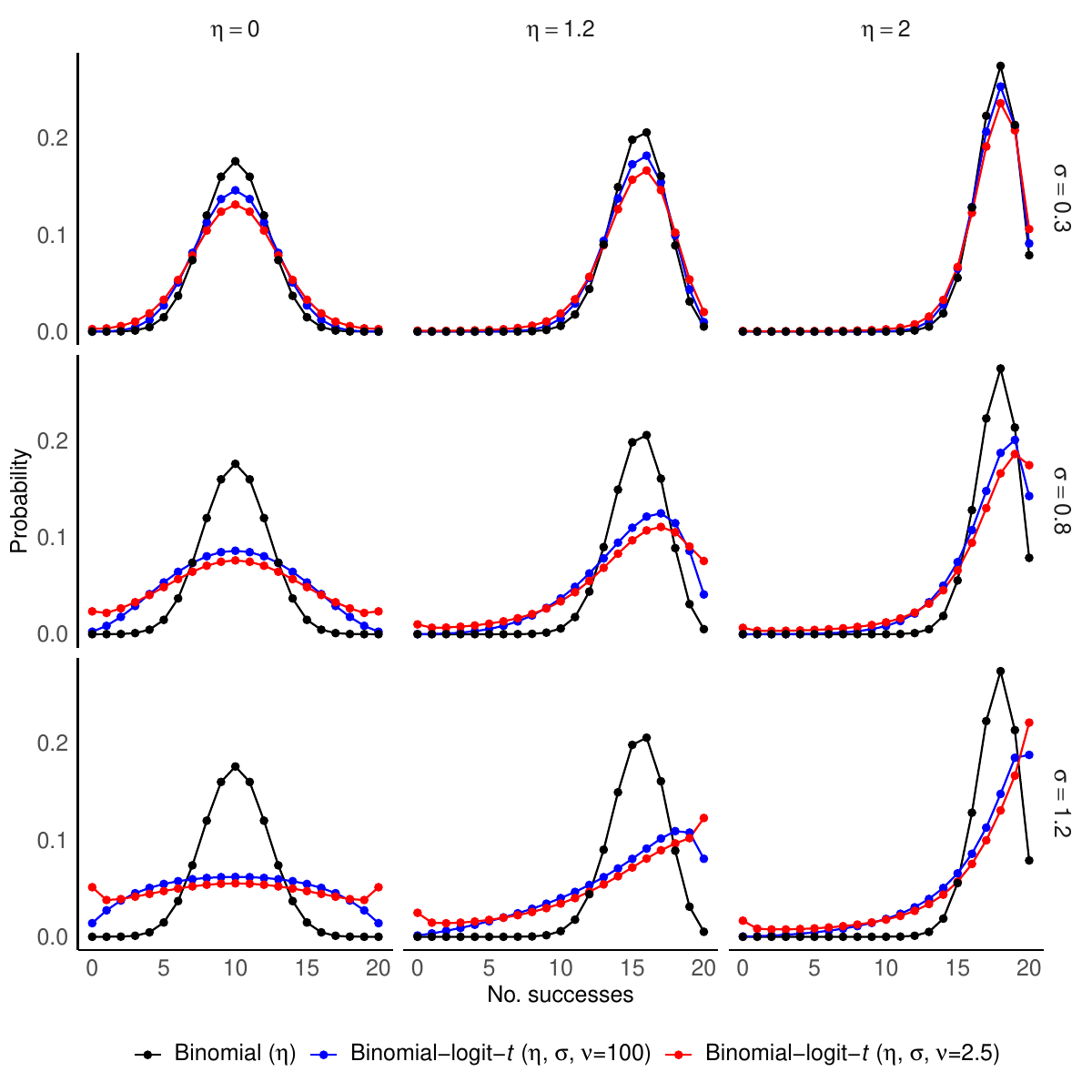}
        \caption{Probability mass function of the binomial-logit-$t$ and binomial distributions for varying $\eta$ (location) and $\sigma$ (scale). Black symbols/lines indicate the binomial distribution, which does not account for overdispersion. Blue symbols/lines ($\nu = 100$) and red symbols/lines ($\nu = 2.5$) represent the binomial-logit-$t$ distribution, both illustrating overdispersion with the same location and scale. The red lines show heavier tails due to the lower degrees of freedom, reflecting the impact of outliers. Increases in $\sigma$ from top to bottom panels result in wider probability spreads, and the rightward shift in $\eta$ across columns shifts the peak towards higher successes.}
        \label{fig:LBT_PMF}
    \end{figure}

    \autoref{fig:LBT_PMF} illustrates the PMFs for the binomial-logit-$t$ distribution and the binomial distribution across various parameter configurations, showcasing their roles in determining the location ($\eta$), scale ($\sigma$), and shape ($\nu$) of the distributions. Each subplot displays the effects of different combinations of $\eta$ and $\sigma$ on the outcome distribution. The binomial distribution, which does not account for overdispersion, is included for comparison. The binomial-logit-$t$ distributions illustrate overdispersion with the same location and scale but differ in tail heaviness due to varying degrees of freedom ($\nu$). The heavier tails of the distributions with lower degrees of freedom reflect the impact of outliers, illustrating the distribution's flexibility in adjusting to various data characteristics, such as overdispersion and the presence of outliers.

    \section{Mixed-effects models for bounded counts} \label{sec:MIXED_MODELS}

    This section formulates the robust mixed-effects regression model for bounded counts, specifically employing the binomial-logit-$t$ model. This model is expected to provide robustness to outliers and accommodate overdispersion. It is compared with its less robust competitors: the beta-binomial and binomial-logit-normal models, which can accommodate overdispersion but are less effective in handling outliers. Additional comparisons include the conventional binomial model, which neither accounts for overdispersion nor provides robustness against outliers. These models aim to model the proportion of successes in bounded counts as a function of covariates.

    In the following subsections, we present a series of mixed-effects models for bounded counts, starting with the simplest and least robust models and progressively introducing more complex assumptions. \autoref{tab:MODEL_CHARACTER} summarizes the key characteristics of each model, including their ability to accommodate overdispersion and outliers and their capacity to model the mean and pseudo-median response proportions.

    \setlength{\tabcolsep}{0.08cm}
    \begin{table}[t!]
    \centering
    \footnotesize
    \caption{Comparison of mixed-effects models for bounded counts. ``Integration required" refers to whether numerical integration is needed to obtain the mean or median (pseudo-median) on the original scale.} \label{tab:MODEL_CHARACTER}
    
    \begin{tabular}{llllll}
    \hline
     & \multicolumn{2}{c}{\bf Handles} & & \multicolumn{2}{c}{\bf Integration required} \\
    \cline{2-3} \cline{5-6}
    \textbf{Mixed model} & \textbf{Overdispersion} & \textbf{Outliers} & 
      & \textbf{Mean} & \textbf{Median (pseudo-median)} \\ 
    \hline
    Binomial             
      & No  & No  
      & & No (closed-form) 
      & No (same as mean) \\
    
    Beta-binomial        
      & Yes & No  
      & & No (closed-form) 
      & Yes (no closed-form) \\
    
    Binomial-logit-normal
      & Yes & No  
      & & Yes (numerical) 
      & No (closed-form) \\
    
    Binomial-logit-$t$   
      & Yes & Yes 
      & & Yes (numerical) 
      & No (closed-form) \\
    \hline
    \end{tabular}
    \end{table}

    Suppose $y_{ij}$ is the bounded count outcome observed from $m_{ij}$ number of trials (or the total count capacity) for cluster $i = 1, \ldots, I$ and observation $j = 1, \ldots, J_i$. Let $\bm{\beta}$ and $\bm{u}_i$ denote fixed and cluster-specific random effects vectors, and $\bm{x}_{ij}$ and $\bm{z}_{ij}$ the covariate vectors, respectively. Assume the $\bm{u}_i$ follow a multivariate normal distribution with mean $\bm{0}$ and a $d$-dimensional unstructured covariance matrix $\bm{\Sigma}$, such that $\bm{u}_i \left| \bm{\Sigma} \right. \sim \mathcal{N}_d(\bm{0}, \bm{\Sigma})$.
    
    Details on the Bayesian specification of the candidate models are presented in \autoref{sec:BAYESSPEC} of the supplementary material. We employ uninformative priors throughout our analysis, including the matrix-generalized half-$t$ (MGH-$t$) prior for $\bm{\Sigma}$ as suggested by Huang and Wand \citep{HUANG2013} and a vague prior for the degrees of freedom parameter $\nu$ following Simpson et al. \citep{SIMPSON2017}.

    \subsection{Binomial model} \label{sec:BINMODEL}

    The conventional binomial model represents scenarios where the probability of success is consistent across trials without accommodating overdispersion. The PMF of the binomial model for the bounded count outcome $y_{ij}$ with $m_{ij}$ trials is given by
    \begin{equation} \label{eq:BINMODEL}
        P\left(y_{ij} \left| \bm{\beta}, \bm{u}_i \right.\right) = \binom{m_{ij}}{y_{ij}} p_{ij}^{y_{ij}} \left(1-p_{ij}\right)^{m_{ij} - y_{ij}},
    \end{equation}
    where $p_{ij}$ is the probability of success, modeled using the logistic regression link function
    \begin{equation} \label{eq:BINLINK}
        \eta_{ij} = \text{logit}\left(p_{ij}\right) = \bm{x}_{ij}^\prime \bm{\beta} + \bm{z}_{ij}^\prime \bm{u}_i.
    \end{equation}
    \begin{equation} \label{eq:BINMODELEXPU0}
        \mathbb{E}\left[y_{ij} \left| \bm{u}_i = \bm{0} \right.\right] = m_{ij} \frac{1}{1 + e^{-\bm{x}_{ij}^\prime \bm{\beta}}}.
    \end{equation}
    Because there is no mixture component, the (pseudo-)median of the binomial distribution equals its mean, both determined by $p_{ij}$.

    \subsection{Beta-binomial model} \label{sec:BBMODEL}

    The beta-binomial model incorporates a mixture of binomial and beta distributions, allowing for variability in the probability of success across trials, effectively addressing overdispersion. The PMF of the beta-binomial model for the bounded count outcome $y_{ij}$ with $m_{ij}$ trials is expressed as
    \begin{equation}
        P\left(y_{ij} \left| \bm{\beta}, \bm{u}_i, \delta \right.\right) = \int_0^1 \binom{m_{ij}}{y_{ij}} p_{ij}^{y_{ij}} (1-p_{ij})^{m_{ij} - y_{ij}} \text{Beta}\left(p_{ij} \left| \theta_{ij}, \gamma_{ij}\right.\right) \mathrm{d}p_{ij},
    \end{equation}
    where $\text{Beta}\left(p_{ij} \left| \theta_{ij}, \gamma_{ij}\right.\right)$ denotes the PDF of the beta distribution. The shape parameters $\theta_{ij} = p_{ij} \delta$ and $\gamma_{ij} = \left(1 - p_{ij}\right) \delta$ are derived from the logistic regression of the probability of success $p_{ij}$, as shown in Equation~(\ref{eq:BINLINK}).

    Given the fixed effects framework where random effects are set to zero for interpretation ($\bm{u}_i = \bm{0}$), the expected value of the response $y_{ij}$ is similarly given by Equation~(\ref{eq:BINMODELEXPU0}).

    This equation straightforwardly quantifies the expected number of successes without additional integration, reflecting the direct impact of covariates adjusted for overdispersion through the beta-binomial formulation.

    The pseudo-median of the beta-binomial distribution is not available in closed form and requires numerical methods to estimate.

    \subsection{Binomial-logit-normal model} \label{sec:BLNMODEL}

    The binomial-logit-normal model is a mixed-effects model that accommodates overdispersion but is not robust to outliers. The PMF of the binomial-logit-normal model for a given bounded count outcome $y_{ij}$ with $m_{ij}$ trials is given by
    \begin{equation}
        P\left(y_{ij} \left| \bm{\beta}, \bm{u}_i, \sigma^2 \right.\right) = \int_{-\infty}^{\infty} \binom{m_{ij}}{y_{ij}} \left(\frac{1}{1 + e^{-\xi_{ij}}}\right)^{y_{ij}} \left(1 - \frac{1}{1 + e^{-\xi_{ij}}}\right)^{m_{ij} - y_{ij}} \mathcal{N}\left(\xi_{ij} \left| \eta_{ij}, \sigma^2\right.\right) \, \mathrm{d}\xi_{ij},
    \end{equation}
    where $\eta_{ij} = \bm{x}_{ij}^\prime \bm{\beta} + \bm{z}_{ij}^\prime \bm{u}_i$ represents the mean of the normal distribution, and $\xi_{ij}$ are the nuisance parameters that are integrated out. The term $\mathcal{N}\left(\xi_{ij} \left| \eta_{ij}, \sigma^2\right.\right)$ denotes the PDF of the normal distribution.

    In order to interpret the fixed effects within the context of an ``average" cluster, where random effects are considered to be zero, the fixed effects ($\bm{\beta}$) are assessed by setting $\bm{u}_i = \bm{0}$. This adjustment simplifies $\eta_{ij}$ to $\bm{x}_{ij}^\prime \bm{\beta}$, allowing interpretation focused solely on the covariates' direct impacts without the variability introduced by random effects, centering on the expected (average) response:
    \begin{equation} \label{eq:UNCOND_EXP_NORM}
        \mathbb{E}\left[y_{ij} \left| \bm{u}_i = \bm{0} \right.\right] = \int_{-\infty}^{\infty} m_{ij} \frac{1}{1 + e^{-\xi_{ij}}} \mathcal{N}\left(\xi_{ij} \left| \eta_{ij}=\bm{x}_{ij}^\prime \bm{\beta}, \sigma^2 \right.\right) \mathrm{d}\xi_{ij}.
    \end{equation}
    This integration process computes the expected value of the response variable, conditional on the random effects being zero, while integrating out the nuisance parameters $\xi_{ij}$.

    The pseudo-median of the binomial-logit-normal model is directly linked to the linear predictor without the need for integration. Specifically, the pseudo-median value is given by
    \begin{equation} \label{eq:MEDIAN_BLN_MODEL}
        \widetilde{M}\left[y_{ij} \left| \bm{u}_i = \bm{0} \right.\right] = m_{ij} \frac{1}{1 + e^{-\bm{x}_{ij}^\prime \bm{\beta}}}.
    \end{equation}
    The model can be used to predict outcomes by setting covariate values in $\bm{x}_{ij}$ for specific scenarios, such as estimating probabilities for a treatment group at a given time point. For mean-based predictions, this requires integrating the latent variables $\xi_{ij}$, while median-based predictions can be directly obtained using the simplified expression in Equation~(\ref{eq:MEDIAN_BLN_MODEL}) without the need for integration.

    As a special case, when $\sigma^2$ approaches zero, the model simplifies to the conventional binomial distribution. In this scenario, the expected value calculation, $\mathbb{E}\left[y_{ij} \left| \bm{u}_i = \bm{0} \right.\right]$, reduces to the expression for the pseudo-median in Equation~(\ref{eq:MEDIAN_BLN_MODEL}), similar to the expected value in standard logistic regression models.
    
    \subsection{Binomial-logit-\textbf{\textit{t}} model} \label{sec:BLTMODEL}

    The binomial-logit-$t$ model, based on the binomial-logit-$t$ distribution introduced in \autoref{sec:BLTDIST}, is a mixed-effects model designed to accommodate both overdispersion and heavy tails in the data. The PMF of the binomial-logit-$t$ model for a given bounded count outcome $y_{ij}$ with $m_{ij}$ trials is given by
    \begin{equation} \label{eq:BLTMODEL}
        P\left(y_{ij} \left| \bm{\beta}, \bm{u}_i, \sigma^2, \nu\right.\right) = \int_{-\infty}^{\infty} \binom{m_{ij}}{y_{ij}} \left(\frac{1}{1 + e^{-\xi_{ij}}}\right)^{y_{ij}} \left(1 - \frac{1}{1 + e^{-\xi_{ij}}}\right)^{m_{ij} - y_{ij}} t\left(\xi_{ij} \left| \eta_{ij}, \sigma^2, \nu\right.\right) \, \mathrm{d}\xi_{ij},
    \end{equation}
    where $\eta_{ij} = \bm{x}_{ij}^\prime \bm{\beta} + \bm{z}_{ij}^\prime \bm{u}_i$, and $\xi_{ij}$ are the nuisance parameters that are integrated out.

    As with the binomial-logit-normal model, setting the random effects to zero ($\bm{u}_i = \bm{0}$) simplifies $\eta_{ij}$ to $\bm{x}_{ij}^\prime \bm{\beta}$. The expected value of $y_{ij}$ is then calculated as
    \begin{equation} \label{eq:UNCOND_MEAN_BLT}
        \mathbb{E}\left[y_{ij} \left| \bm{u}_i = \bm{0} \right.\right] = \int_{-\infty}^{\infty} m_{ij} \frac{1}{1 + e^{-\xi_{ij}}} t\left(\xi_{ij} \left| \eta_{ij}=\bm{x}_{ij}^\prime \bm{\beta}, \sigma^2, \nu \right.\right) \mathrm{d}\xi_{ij}.
    \end{equation}
    Similarly, the pseudo-median value of $y_{ij}$ can be directly linked to the linear predictor without the need for integration as in Equation~(\ref{eq:MEDIAN_BLN_MODEL}).

    When $\sigma^2$ approaches zero, and $\nu$ tends to infinity, $\xi_{ij}$ converges to $\eta_{ij}$, reducing the model to the conventional binomial distribution. In this case, the expected value in Equation~(\ref{eq:UNCOND_MEAN_BLT}) simplifies to Equation~(\ref{eq:MEDIAN_BLN_MODEL}).

    \section{Model comparison and adequacy}\label{sec:MOD_COMP_ADE}

    We evaluate model performance using three complementary criteria.

    First, we apply the WAIC \citep{watanabe2010asymptotic}, calculated using the \verb|loo| package \citep{vehtari2024package}. WAIC balances model fit and complexity via
    \begin{equation}
      \text{WAIC} = \operatorname{lppd} - p_{\text{WAIC}},
    \end{equation}
    where $\operatorname{lppd}$ is the log pointwise predictive density and $p_{\text{WAIC}}$ is the effective number of parameters. Lower WAIC values indicate better expected out-of-sample performance.

    Second, we approximate the K-L divergence between each model's posterior predictive distribution and the empirical distribution using the leave-one-cluster-out importance-sampling estimator described in \autoref{sec:KL_METHOD} \citep{WANG2016, TOMAZELLA2021, LESAFFRE2012}. Large K-L values flag influential observations and possible model misspecification.

    Third, we assess goodness of fit using posterior predictive residual diagnostics from the \verb|DHARMa| package \citep{hartig2022package}. We report Kolmogorov-Smirnov $p$-values for uniformity, parametric bootstrap $p$-values for overdispersion, and binomial $p$-values for outliers; $p$-values below 0.05 signal a lack of fit.

    \section{Data analysis} \label{sec:APPLICATION}

    \subsection{Model implementation}

    We applied the mixed-effects binomial, beta-binomial, binomial-logit-normal, and binomial-logit-$t$ models, as described in \autoref{sec:MIXED_MODELS}, to analyze the medication adherence data detailed in \autoref{sec:MOTIVATE_DATA}.

    The binomial-logit-$t$ model was selected for its robustness to both overdispersion and outliers, likely yielding the best fit given the outliers present in the adherence data. The binomial-logit-normal and beta-binomial models, which address overdispersion but not outliers, along with the conventional binomial model that accommodates neither, are included as comparative baselines to demonstrate the superiority of the binomial-logit-$t$ model.

    We modeled the mean proportion under all four models and the pseudo-median with the binomial-logit-normal and binomial-logit-$t$ models to compare model fits for both central tendency measures.
    
    For each model, $y_{ij}$ represents the number of days the medication package was opened within the recent observation period of $m_{ij}$ days for each patient $i = 1, \ldots, 392$ across $j = 1, \ldots, J_i$ observations. Here, $y_{ij}$ and $m_{ij}$ together quantify adherence, capturing both the actual and potential medication-taking events for each observation period.

    In the linear predictor for all models, $\eta_{ij}=\bm{x}_{ij}^\prime \bm{\beta} + \bm{z}_{ij}^\prime \bm{u}_i$, the vector $\bm{x}_{ij}$ includes covariates for the fixed effects $\bm{\beta}=\left(\beta_0, \beta_{\text{group}}, \beta_{\text{time}}, \beta_{\text{tx}}, \beta_{\text{age}}\right)^\prime$, such as the intercept, treatment group, time, group-time interaction, and patient age. Conversely, $\bm{z}_{ij}$ contains covariates for the random effects $\bm{u}_i=\left(a_0, a_1\right)^\prime$, including the intercept and time:
    \begin{equation}
        \bm{x}_{ij}=\left(1, \text{group}_i, \text{time}_{ij}, \text{group}_i \times \text{time}_{ij}, \text{age}_i\right)^\prime,
    \end{equation}
    \begin{equation}
        \bm{z}_{ij}=\left(1, \text{time}_{ij}\right)^\prime.
    \end{equation}
    Here, $\text{group}_i=0$ indicates that patient $i$ belongs to the control group, while $\text{group}_i=1$ indicates assignment to the intervention group. The variable $\text{time}_{ij}$ represents the month of the $j^\text{th}$ observation for patient $i$, and $\text{age}_i$ denotes the age of the patient (in years).

    Furthermore, $\bm{\Sigma}$ is the unstructured covariance matrix of $\bm{u}_i$, which is a $2 \times 2$ matrix defined as
    \begin{equation}
        \bm{\Sigma} = \begin{bmatrix}
        \sigma^2_{a_0} & \sigma_{a_0, a_1} \\
        \sigma_{a_0, a_1} & \sigma^2_{a_1} \end{bmatrix},
    \end{equation}
    where $\sigma^2_{a_0}$ and $\sigma^2_{a_1}$ are the variances of the random effects for the intercept and time, respectively, and $\sigma_{a_0, a_1}$ represents the covariance between them. In the posterior summaries presented, $\bm{\Sigma}^{-1}$, the inverse of the unstructured covariance matrix $\bm{\Sigma}$, is used.
    
    The models were fitted using \verb|JAGS| via the \verb|R| package \verb|runjags| \citep{DENWOOD2016}. Our study generated 27~500 samples from each model's joint posterior distribution using four parallel chains. We discarded the first 2~500 samples from each chain as burn-in. The adequacy of sample convergence was assessed using trace plots and the Brooks-Gelman-Rubin diagnostic \citep{BROOKS1998}. We applied a thinning factor of 50 to mitigate sample autocorrelation, resulting in a total of 2~000 samples (500 per chain) per model parameter for analysis. The model parameters' posterior distributions were summarized using point and interval estimates. The median of the posterior samples for each parameter was reported as the posterior estimate, and the 95\% highest posterior density (HPD) intervals were calculated to represent the shortest intervals containing 95\% of the posterior samples.

    We estimated the adherence proportions over time for each intervention group under each regression model. For each posterior sample, adherence proportions were calculated for a series of time points across the different intervention groups. Specifically, we computed the predicted probabilities of adherence for each combination of time (from months~0 to 10) and intervention group (intervention and non-intervention), setting the random effects to zero to represent the average patient and using the average age as a covariate. For the binomial-logit-normal and binomial-logit-$t$ models, the calculated mean probabilities are not conditional on the distributions' latent (or nuisance) parameters $\xi_{ij}$, aligning with Equations~(\ref{eq:UNCOND_MEAN_BLT}) and (\ref{eq:UNCOND_EXP_NORM}), where expected values are computed independent of $\xi_{ij}$. Additionally, we computed the closed-form pseudo-median probabilities for the binomial-logit-normal and binomial-logit-$t$ models directly from the linear predictors (Equation~(\ref{eq:MEDIAN_BLN_MODEL})), avoiding the need for integration and thereby staying close to the discrete median.

    We summarized the posterior results by calculating the median of the predicted mean probabilities for each time point and intervention group, serving as estimates for mean adherence proportions. For the binomial-logit-normal and binomial-logit-$t$ models, we also calculated the median of predicted median probabilities to estimate pseudo-median adherence proportions. Additionally, 95\% HPD intervals were computed to quantify the uncertainty of each estimate.

    We calculated WAICs and K-L divergence estimates and conducted residual checks to compare models, identify outliers, and determine the most suitable model.

    Example code to reproduce the analysis and generate similar output can be found in the GitHub repository at \href{https://github.com/DABURGER1/Robust-Logistic-Regression}{https://github.com/DABURGER1/Robust-Logistic-Regression}. This repository includes a detailed \verb|R| Markdown file with step-by-step examples based on simulated data to demonstrate our model's application and facilitate ease of use.

    All models were run on a 4-core Intel i5-1145G7 laptop (16~GB RAM). Convergence times were modest: binomial $\approx$~5~min, beta-binomial $\approx$~120~min, binomial-logit-normal $\approx$~8~min, and binomial-logit-$t$ $\approx$~20~min. Thus, the added robustness of the $t$ model incurs only a moderate computational cost.

    On a separate simulated dataset mirroring the adherence-study design, we also fitted the binomial-logit-$t$ model in \verb|Stan| using its default No-U-Turn Sampler (NUTS) \citep{Carpenter2017}. Posterior summaries were similar to the \verb|JAGS| results, and total run-time was of the same order. We, therefore, retain \verb|JAGS| as the primary engine while making the \verb|Stan| code available in the project's GitHub repository for readers who prefer Hamiltonian Monte Carlo sampling.

    \subsection{Regression fits and model comparison}

    Web \autoref{tab:posterior_summary} in \autoref{sec:ADDIT_OUTPUT} of the supplementary material presents the posterior estimates and 95\% HPD intervals for the model parameters across all four models, including the WAICs.

    The 95\% HPD interval for $\nu$ of the binomial-logit-$t$ model is less than 2, indicating a significant presence of outliers in the data, as lower values of $\nu$ suggest heavier tails in the distribution.

    The WAICs ranked the robust binomial-logit-$t$ model as the most suitable, followed by the binomial-logit-normal model, the beta-binomial model, and lastly, the conventional binomial model, demonstrating their respective adequacy in modeling the data.

\begin{figure}[b!]
        \centering
        \includegraphics[trim=0.0cm 0.0cm 0.0cm 0.0cm, clip=true, scale=0.85]{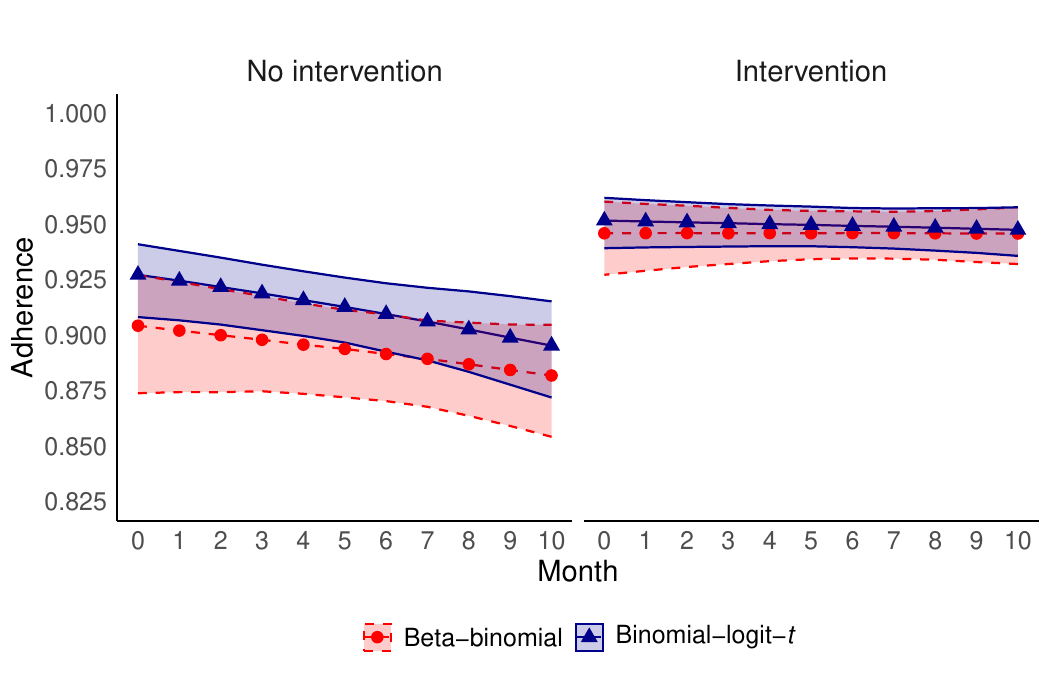}
        \caption{Mean adherence proportion estimates and 95\% highest posterior density (HPD) intervals under the beta-binomial and binomial-logit-$t$ models over a 10-month period. The left panel shows non-intervention patients, while the right panel shows intervention patients. Red dashed lines with dot markers represent the beta-binomial model, and blue solid lines with triangle markers represent the binomial-logit-$t$ model. The 95\% HPD intervals for the non-intervention group are wider for the beta-binomial model, indicating greater uncertainty in its estimates.}
        \label{fig:BB_LBT_mean_profiles}
    \end{figure}

\begin{figure}[b!]
        \centering
        \includegraphics[trim=0.0cm 0.0cm 0.0cm 0.0cm, clip=true, scale=0.85]{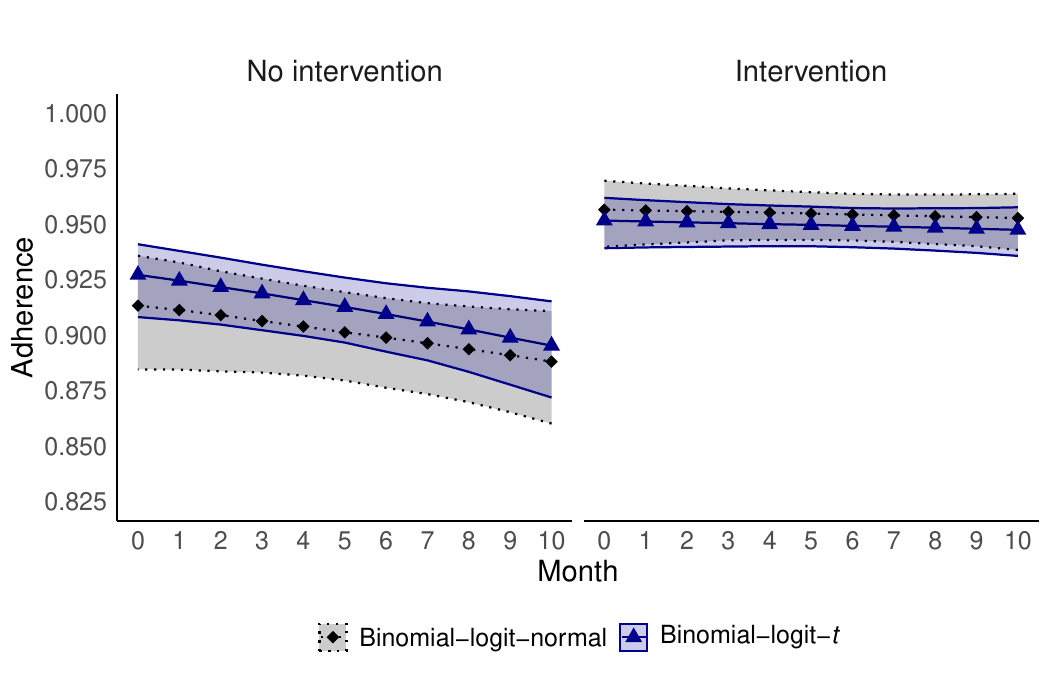}
        \caption{Mean adherence proportion estimates and 95\% highest posterior density (HPD) intervals under the binomial-logit-normal and binomial-logit-$t$ models over a 10-month period. The left panel shows non-intervention patients, while the right panel shows intervention patients. Black dotted lines with diamond markers represent the binomial-logit-normal model, and blue solid lines with triangle markers represent the binomial-logit-$t$ model. The 95\% HPD intervals for the non-intervention group are wider for the binomial-logit-normal model, indicating greater uncertainty in its estimates.}
        \label{fig:LBN_LBT_mean_profiles}
    \end{figure}

\begin{figure}[b!]
        \centering
        \includegraphics[trim=0.0cm 0.0cm 0.0cm 0.0cm, clip=true, scale=0.85]{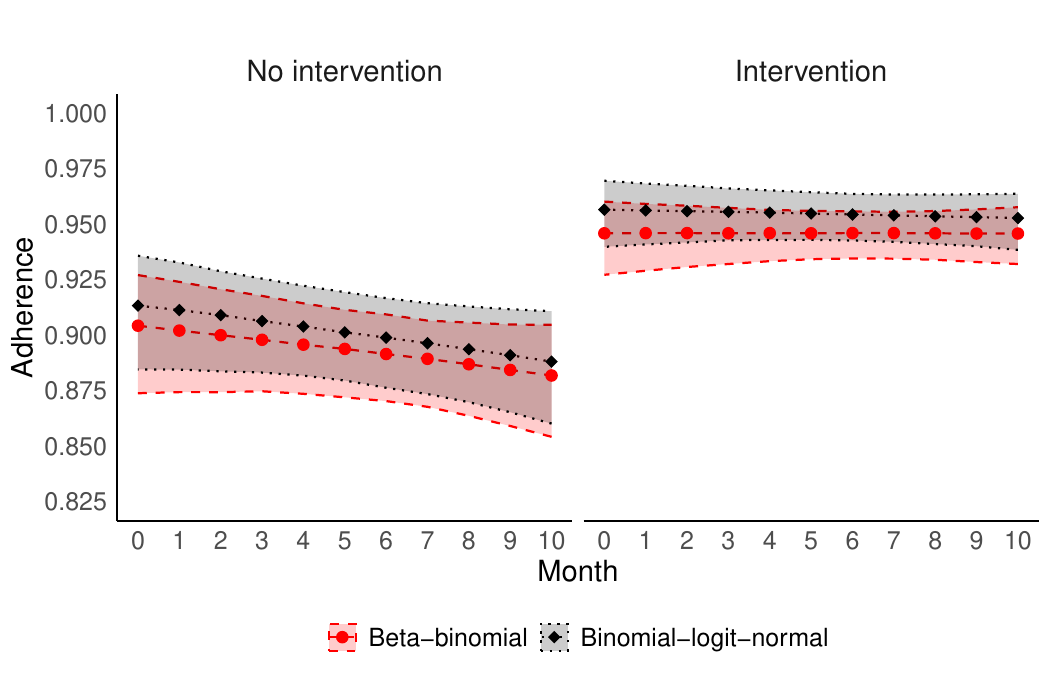}
        \caption{Mean adherence proportion estimates and 95\% highest posterior density (HPD) intervals under the beta-binomial and binomial-logit-normal models over a 10-month period. The left panel shows non-intervention patients, while the right panel shows intervention patients. Red dashed lines with dot markers represent the beta-binomial model, and black dotted lines with diamond markers represent the binomial-logit-normal model. The estimated adherence proportions and 95\% HPD intervals are similar between the models.}
        \label{fig:BB_LBN_mean_profiles}
    \end{figure}

\begin{figure}[b!]
        \centering
        \includegraphics[trim=0.0cm 0.0cm 0.0cm 0.0cm, clip=true, scale=0.85]{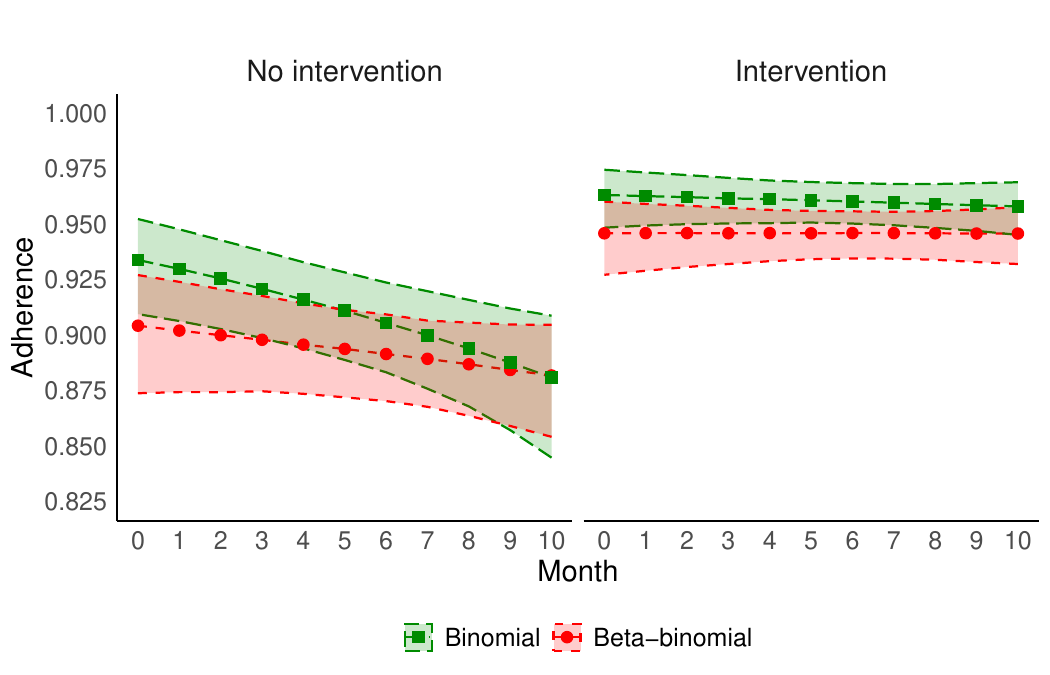}
        \caption{Mean adherence proportion estimates and 95\% highest posterior density intervals under the binomial and beta-binomial models over a 10-month period. The left panel shows non-intervention patients, while the right panel shows intervention patients. Green long-dashed lines with square markers represent the binomial model, and red dashed lines with dot markers represent the beta-binomial model.}
        \label{fig:Bin_BB_mean_profiles}
    \end{figure}

    We contrast the mean adherence proportion estimates and 95\% HPD intervals between the intervention and non-intervention groups over a 10-month period across different models pairwise. We begin by comparing the robust overdispersed model (binomial-logit-$t$) with the overdispersion-only models (beta-binomial and binomial-logit-normal) in {\color{red} Figures}~\ref{fig:BB_LBT_mean_profiles} and \ref{fig:LBN_LBT_mean_profiles}. Next, we evaluate the two overdispersion-only models against each other in \autoref{fig:BB_LBN_mean_profiles}. Finally, we compare the commonly used conventional models, contrasting the beta-binomial model with the standard binomial model in \autoref{fig:Bin_BB_mean_profiles}. {\color{red} Web} \autoref{fig:All_models_mean_profiles} in \autoref{sec:ADDIT_OUTPUT} of the supplementary material presents the adherence proportion estimates under all four models combined.

\begin{figure}[b!]
        \centering
        \includegraphics[trim=0.0cm 0.0cm 0.0cm 0.0cm, clip=true, scale=0.85]{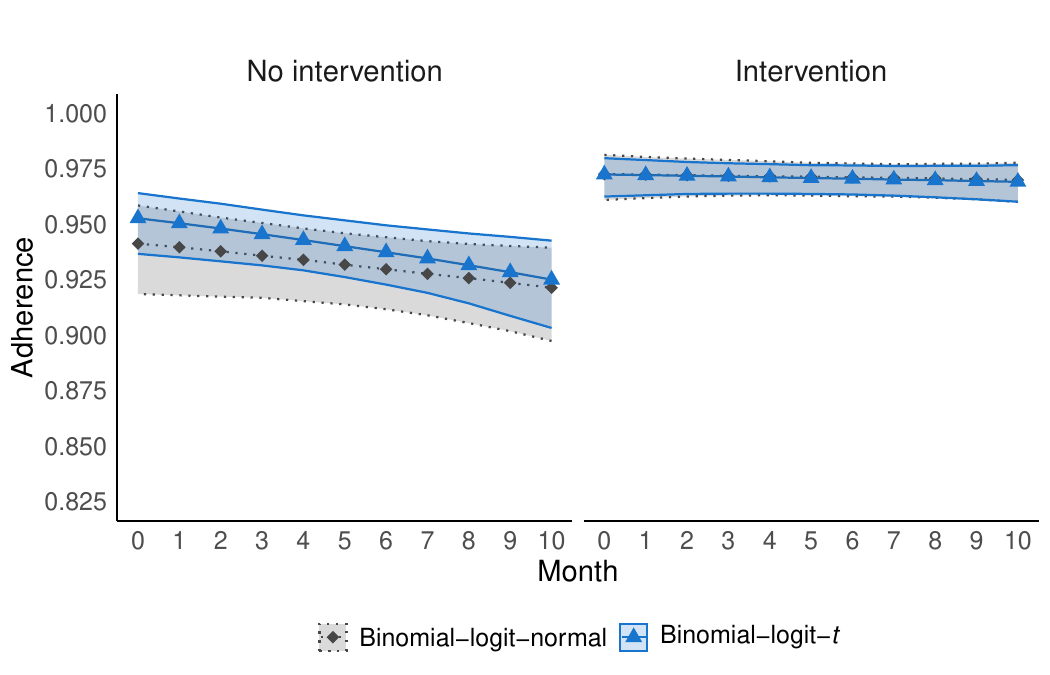}
        \caption{Pseudo-median adherence proportion estimates and 95\% highest posterior density (HPD) intervals under the binomial-logit-normal and binomial-logit-$t$ models over a 10-month period. The left panel shows non-intervention patients, while the right panel shows intervention patients. Gray dotted lines with diamond markers represent the binomial-logit-normal model, and dodgerblue solid lines with triangle markers represent the binomial-logit-$t$ model. The results show that the 95\% HPD intervals for the non-intervention group are slightly wider for the binomial-logit-normal model. For the intervention group, the results are similar between the two models.}
        \label{fig:LBN_LBT_median_profiles}
    \end{figure}

    In addition to evaluating mean adherence, we assessed pseudo-median adherence proportions. This analysis examines the pseudo-median adherence estimates between the binomial-logit-normal and binomial-logit-$t$ models. The pseudo-median adherence proportions and their corresponding 95\% HPD intervals for the intervention and non-intervention groups over the 10-month period are presented in \autoref{fig:LBN_LBT_median_profiles}.
 
    The results from {\color{red} Figures}~\ref{fig:BB_LBT_mean_profiles} to \ref{fig:LBN_LBT_median_profiles} and {\color{red} Web} \autoref{fig:All_models_mean_profiles} reveal the following:
    \begin{itemize}
        \item The mean adherence proportions estimated by all four models are broadly comparable.
        \item The non-robust overdispersed models, specifically the beta-binomial and binomial-logit-normal, exhibit wider 95\% HPD intervals compared to the robust overdispersed model (binomial-logit-$t$). This suggests greater uncertainty in their estimates, likely due to less effective handling of outliers present in the adherence data.
        \item The mean adherence estimates and associated 95\% HPD intervals from the beta-binomial and binomial-logit-normal models are notably similar, reflecting their comparable handling of overdispersion. However, they lack specific mechanisms to address outlier effects robustly.
        \item The mean adherence estimates between the beta-binomial and standard binomial models are comparable, with any differences potentially attributable to the beta-binomial model's capacity to accommodate overdispersion in the data.
        \item The 95\% HPD intervals of the pseudo-median adherence proportions for the non-intervention group are slightly wider under the binomial-logit-normal model compared to the binomial-logit-$t$ model, reflecting greater uncertainty. In contrast, the adherence estimates for the intervention group exhibit similar levels of uncertainty between the two models.
    \end{itemize}

    \subsection{Model adequacy}

\begin{figure}[b!]
        \centering
        \subfloat[Binomial]{\includegraphics[trim = 0cm 0cm 0cm 0cm, clip = true, scale = 0.36] {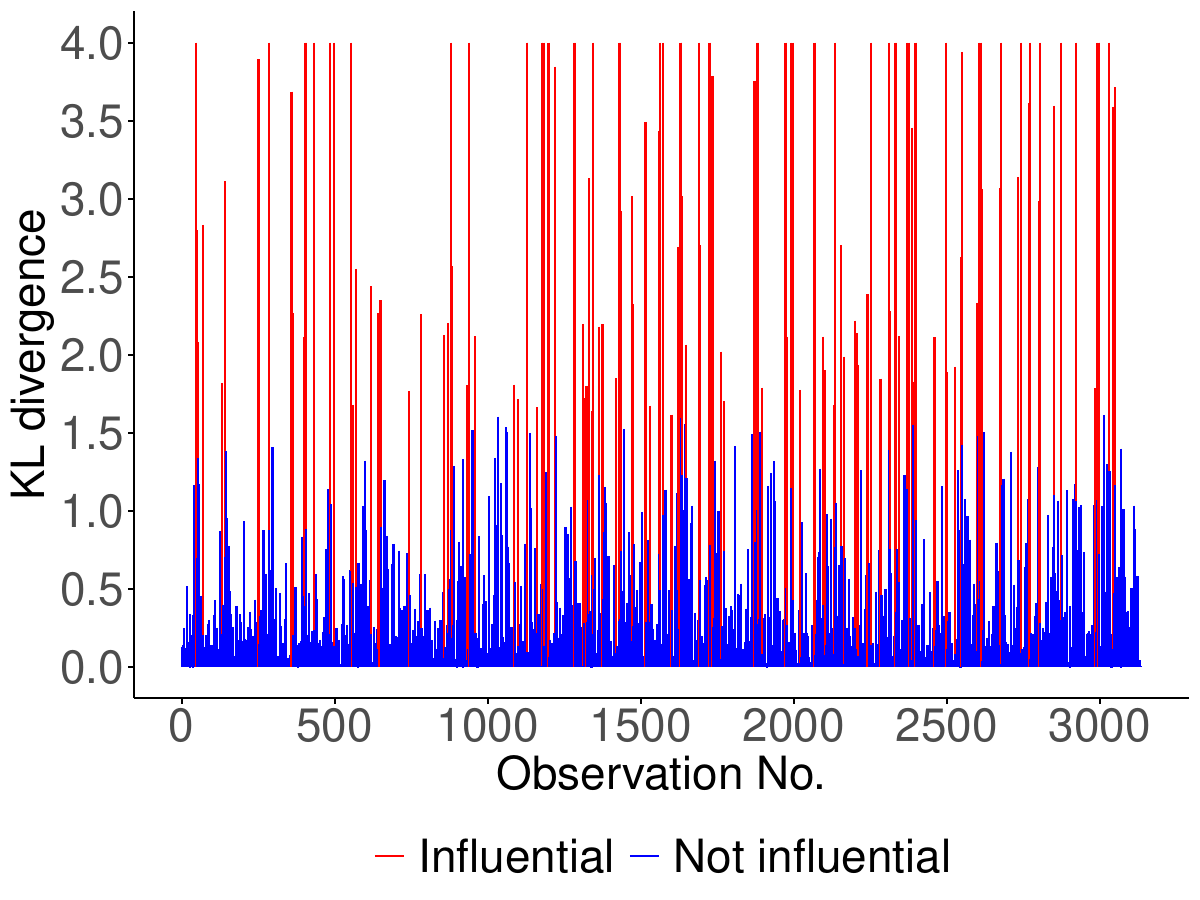}}
        \subfloat[Beta-binomial]{\includegraphics[trim = 0cm 0cm 0cm 0cm, clip = true, scale = 0.36] {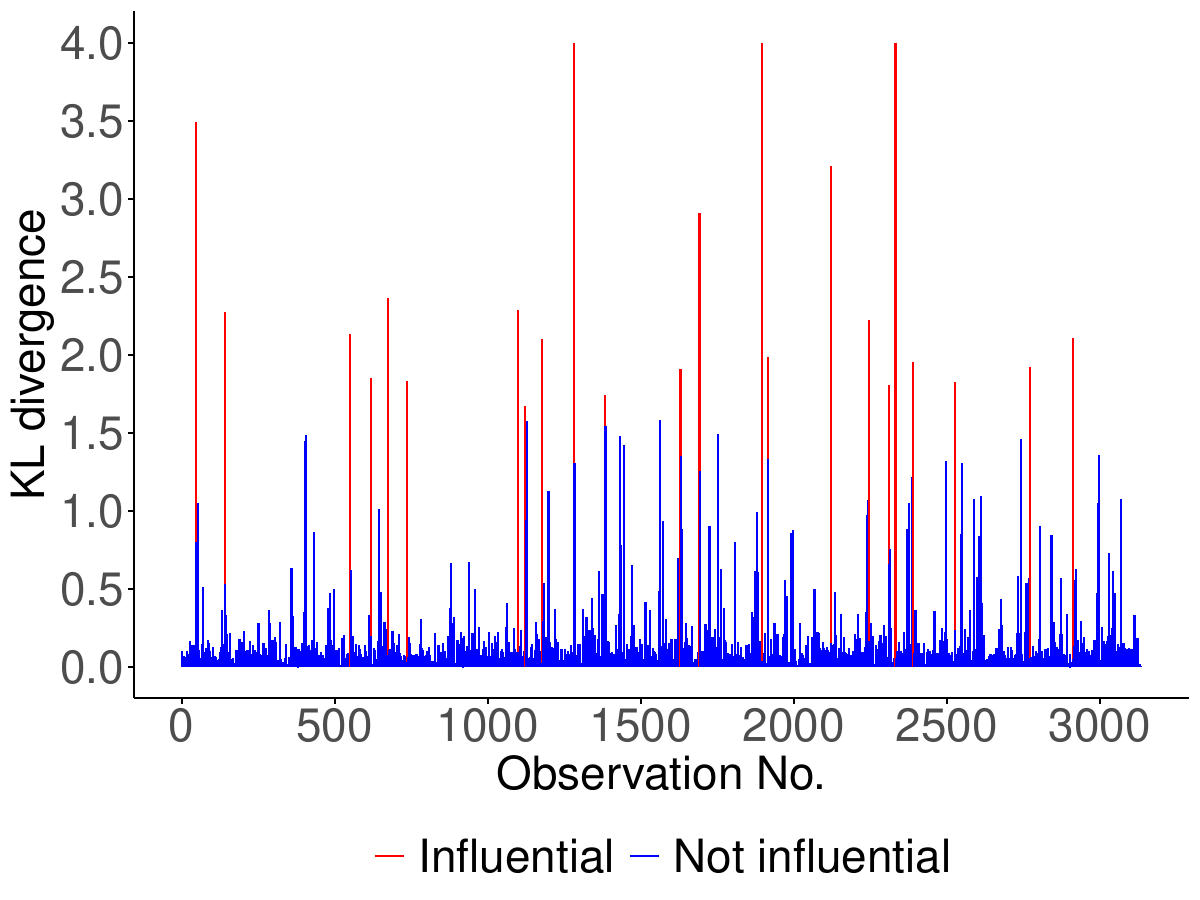}} \\
        \subfloat[Binomial-logit-normal]{\includegraphics[trim = 0cm 0cm 0cm 0cm, clip = true, scale = 0.36] {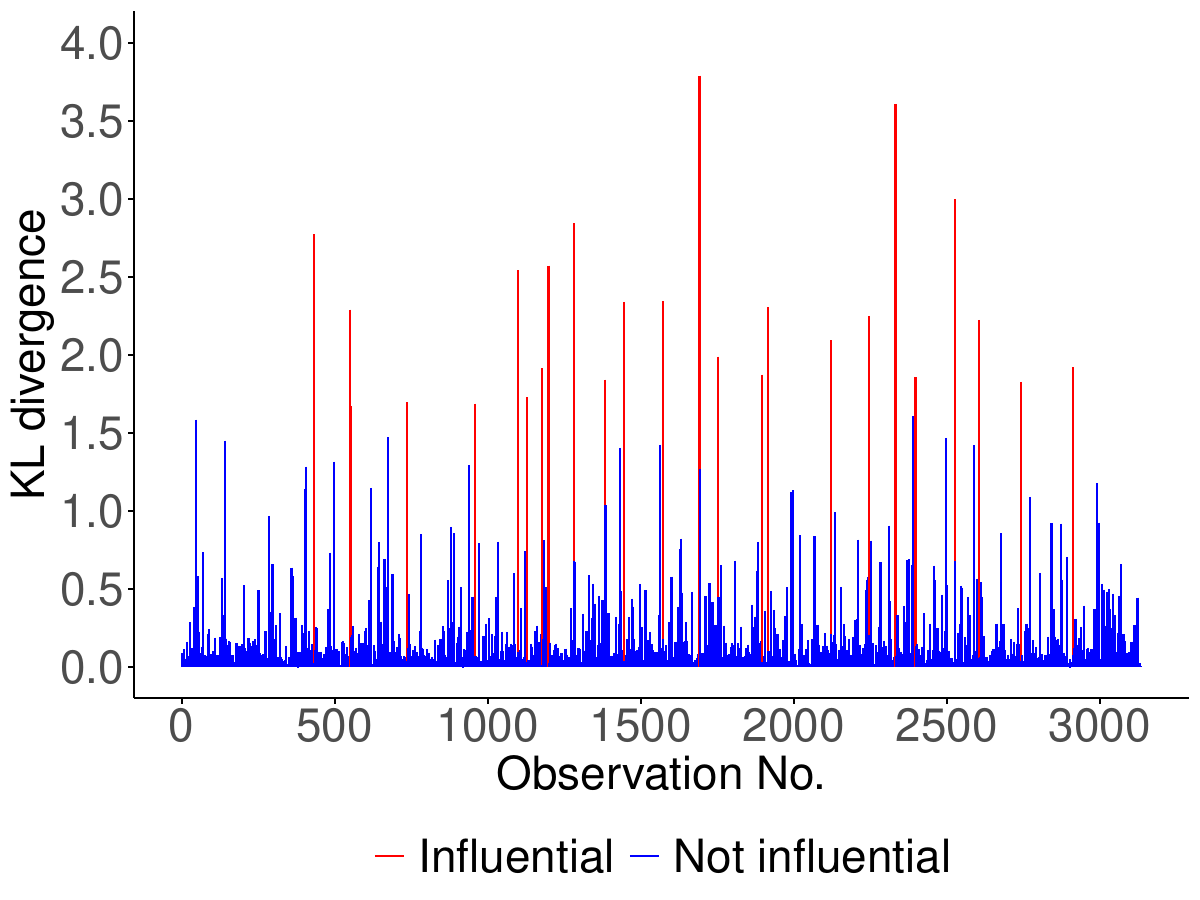}}
        \subfloat[Binomial-logit-$t$]{\includegraphics[trim = 0cm 0cm 0cm 0cm, clip = true, scale = 0.36] {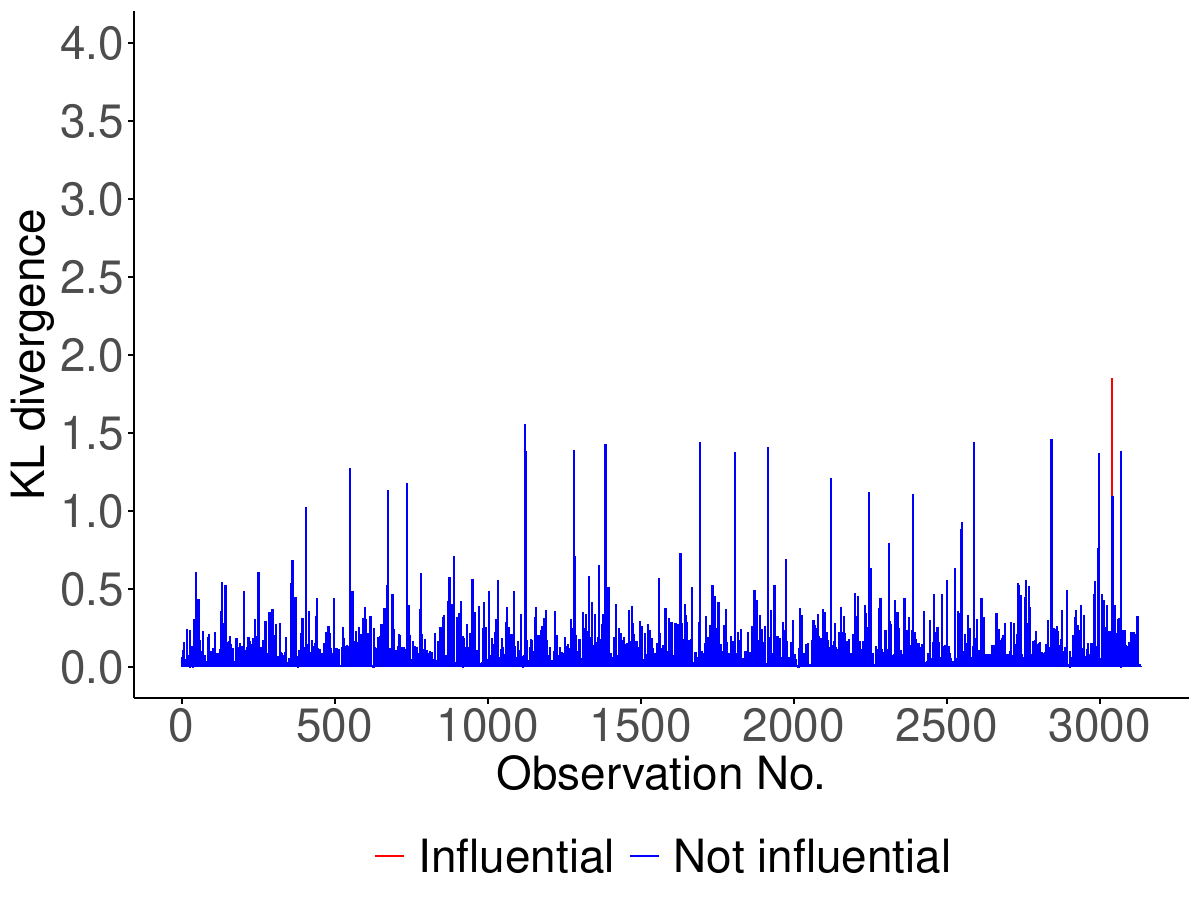}}
        \caption{Kullback-Leibler divergence estimates under the binomial, beta-binomial, binomial-logit-normal, and binomial-logit-$t$ models. Values exceeding 4 are capped. Blue lines represent non-outliers, while red lines indicate influential observations (outliers). The binomial model shows many influential observations, indicating high sensitivity to outliers. The beta-binomial and binomial-logit-normal models show improved robustness, with fewer influential observations. The binomial-logit-$t$ model exhibits the greatest robustness, handling outliers and overdispersion effectively.}
        \label{fig:KL_Estimates}
    \end{figure}

    \autoref{fig:KL_Estimates} presents K-L divergence estimates under the binomial, beta-binomial, binomial-logit-normal, and binomial-logit-$t$ models. The results highlight the robustness of the binomial-logit-$t$ model, which consistently shows lower divergence values, indicating its resilience to outliers. In contrast, the conventional binomial model exhibits significantly higher divergence, suggesting it is substantially affected by influential observations. The beta-binomial and binomial-logit-normal models display moderate divergence levels, indicating some sensitivity to outliers but improved robustness over the standard binomial model.
     
    {\color{red} Web} \autoref{fig:Residual_check} illustrates the results of the residual checks, while \autoref{tab:residual_checks} summarizes the p-values from the uniformity, dispersion, and outlier tests. The residual checks suggest the following:
    \begin{itemize}
        \item The binomial-logit-$t$ model performs the best, with no significant issues in uniformity, dispersion, or outliers.
        \item The binomial-logit-normal, beta-binomial, and binomial models both exhibit significant problems with uniformity and outliers.
    \end{itemize}

    \begin{table}[t!]
    \centering
    \caption{Summary of p-values from residual checks.} \label{tab:residual_checks}
    \begin{tabular}{lccc}
      \hline
    {\bf Model} & {\bf Uniformity} & {\bf Dispersion} & {\bf Outliers} \\ 
      \hline
    Binomial-logit-$t$ & 0.529 & 0.157 & 0.563 \\ 
      Binomial-logit-normal & 0.032 & 0.218 & $<0.001$ \\ 
      Beta-binomial & $<0.001$ & 0.140 & $<0.001$ \\ 
      Binomial & $<0.001$ & 0.151 & $<0.001$ \\ 
       \hline
    \end{tabular}
    \end{table}

    Overall, the binomial-logit-$t$ model emerges as the most reliable choice for analyzing this data.

    \section{Simulation study} \label{sec:SIM_PERF_STUDY}

    In this section, we briefly summarize the simulation study designed to evaluate the performance of the binomial-logit-$t$ model for bounded counts. \autoref{sec:APP_SIM_STUDY} of the supplementary material provides detailed methodology, scenarios, parameters, and results. The primary focus of the simulation study was to assess the performance of the binomial-logit-$t$ model under different conditions of overdispersion and outliers, including a comparison of its robustness against outliers with the binomial-logit-normal model.
    
    For the model performance simulation study, four scenarios were considered, varying in $\sigma$ and $\nu$ values, while other parameters were kept constant to mimic the characteristics of the medication adherence dataset over a 300-day period. For the data contamination simulation study, we selected one scenario, simulated data from the binomial-logit-normal model, and introduced outliers to assess the models' robustness.

    The bias and coverage of parameter estimates were appropriate, indicating that the model has good frequentist properties. The binomial-logit-$t$ model demonstrated greater robustness, as evidenced by lower root mean square error (RMSE) and shorter average 95\% HPD interval lengths compared to the binomial-logit-normal model. For detailed results and further discussion, refer to the supplementary material (\autoref{sec:APP_SIM_STUDY}).

    \section{Discussion} \label{sec:DISCUSSION}

    This study has demonstrated the utility of employing outlier-robust models in analyzing hierarchically structured bounded count data, particularly in contexts where data anomalies can skew traditional analyses. The robust modeling approach adopted here eliminates the need for explicit identification and removal of outliers, thereby preserving the dataset's integrity and ensuring that extreme values do not disproportionately influence the model outcomes.

    A significant advantage of our approach is the availability of a closed-form solution for the pseudo-median of the binomial-logit-$t$ distribution, which contrasts with the model proposed by Bayes et al. \citep{bayes2024robust} that provides a closed-form expression for the mean. This simplification facilitates a more straightforward interpretation of the regression parameters, especially in the context of skewed data, where the (pseudo-)median is often considered a more appropriate measure of central tendency than the mean. Initially, we attempted to implement Bayes et al.'s~\citep{bayes2024robust} model using a Bayesian framework and extend it to a mixed model setting. However, we encountered convergence issues, indicating the potential computational challenges of their approach. Despite these issues, our model shows adequate performance in terms of bias and coverage, demonstrating its effectiveness in analyzing overdispersed and heavy-tailed count data. For the mean proportion predictions, each integration per posterior iteration requires numerical integration to account for the latent variables. Our code runs efficiently by utilizing \verb|R|'s built-in functions to specify the binomial PMF and $t$-distribution's PDF, simplifying the integral calculation.

    A Gibbs implementation using the Pólya-gamma augmentation of Polson et al. \citep{polson2013bayesian} is also possible; it would render the logistic part conditionally Gaussian and could improve mixing, but requires either a custom \verb|JAGS| module or a different engine such as \verb|Stan|.

    The specified multivariate normal distribution for the cluster-specific random effects, $\bm{u}_i$, is not robust to outliers at the cluster level; however, the binomial-logit-$t$ model maintains robustness to outliers on a within-cluster basis. In order to improve robustness against outliers in cluster-specific effects, the model could be extended by adopting a multivariate $t$-distribution for the random effects.
    
    For greater flexibility, while still retaining a closed-form pseudo-median, the latent variable $\xi_{ij}$ can be modeled with heavier-tailed distributions whose median is fixed at $\eta_{ij}$. Symmetric examples include the log-regularly varying scale mixture of normals proposed by Hamura et al. \citep{hamura2022log} and the super-heavy-tailed variance mixture of normals that arises from the whole-robust prior of Gagnon et al. \citep{GAGNON2020}. Because both families are centered at $\eta_{ij}$, the closed-form pseudo-median $\widetilde{M}\left[y_{ij} \left| \bm{u}_i = \bm{0} \right.\right]$ is retained. If directional outliers need to be modeled, one could instead adopt a skew-$t$ or a more general skew-elliptical distribution. These asymmetric laws add tail and shape flexibility, but the median is no longer available in closed form and will generally differ from $\eta_{ij}$; the pseudo-median would therefore have to be computed numerically.

    In our analysis, model comparison is conducted using the conventional WAIC, which conditions on the random effects. While this conditional WAIC provides a useful measure of model fit and complexity, the marginal version of this criterion might give a more appropriate assessment for population-average effects by integrating out the random effects, as discussed by Ariyo et al. \citep{ariyo2020bayesian}. However, calculating the marginal WAIC would involve integrating over the random effects, which could significantly increase computational demands.

    The structural differences between our models, specifically between the binomial and beta-binomial models and the binomial-logit-normal and binomial-logit-$t$ models, influence their comparability when using the WAIC for model assessment. Unlike the binomial and beta-binomial models that directly link the probability of success to the covariates and predict the mean response, the binomial-logit-normal and binomial-logit-$t$ models essentially link the pseudo-median of $Y$ (on the count scale) to the covariates, whereas the binomial/beta-binomial link the mean of $Y$ directly to the covariates. This fundamental difference in the target of estimation adds complexity to the models' structures, which could affect the WAIC's ability to compare model fit and complexity fairly across different model types. This discrepancy is important to consider when interpreting the WAIC results, as it may lead to biased comparisons due to the different targets (mean vs. pseudo-median).

    In conclusion, the binomial-logit-$t$ model presents a robust solution for analyzing bounded count data with overdispersion and outliers. A key advantage of our model is the closed-form expression for the pseudo-median, providing a straightforward interpretation of regression parameters. Although estimating the mean involves numerical integration, we successfully model both the mean and pseudo-median proportions, making our approach versatile in mixed modeling frameworks. The strong simulation performance and demonstrated robustness in real-world applications confirm the practical effectiveness of our model. Future work should focus on optimizing computational methods and extending the model to accommodate a broader range of data structures.

    \section*{Acknowledgements}
    
    We thank AARDEX Group for providing access to the adherence and persistence dataset. This work is based on the research supported by the National Research Foundation (NRF) of South Africa (Grant number 132383). Opinions expressed and conclusions arrived at are those of the authors and are not necessarily to be attributed to the NRF.
    
    \section*{Declaration of conflicting interests}
    
    The authors declared no potential conflicts of interest regarding this article's research, authorship, and/or publication.
    
    \section*{Supplementary material}
    
    Supplementary materials for this article are available online.
    
    Readers interested in the code used for our analysis can find example code in the GitHub repository at \href{https://github.com/DABURGER1/Robust-Logistic-Regression}{https://github.com/DABURGER1/Robust-Logistic-Regression}.

    \section*{Appendix. Pseudo-median vs. exact median}

    We show that the closed-form pseudo-median
    \begin{equation}
        \widehat{y} = g\left(\eta\right) = m_i\text{logit}^{-1}\left(\eta\right)
    \end{equation}
    differs from the discrete median of the binomial-logit-$t$ mixture by strictly less than one count. Define
    \begin{equation}
        k_{i-} = \min \left\{ k_i \in \mathbb{Z} : P\left(Y_i \le k_i\right) \ge 0.5 \right\},
    \end{equation}
    the left-most integer whose cumulative probability reaches 50\%. By definition, $P\left(Y_i \le k_{i-} - 1\right) < 0.5$.

    Since $\eta$ is the median of $\xi_i$ and the logistic map is strictly increasing, $\text{logit}^{-1}\left(\eta\right)$ is the median on the probability scale. However, once the binomial draw is introduced, $Y_i$ can only take integer values, so its median may not equal $\widehat{y}$ exactly. We analyze two possibilities, using the fact that $Y_i$ is stochastically increasing in $\xi_i$: a higher $\xi_i$ induces a larger success probability $p_i$, thus typically increasing $Y_i$.

    \Needspace{4\baselineskip}
    \textbf{Case 1:}
    \begin{equation}
        \widehat{y} = g\left(\eta\right) < k_{i-}.
    \end{equation}
    Then with probability above $0.5$, we must have
    \begin{equation}
        \xi_i < g^{-1}\left(k_{i-}\right),
    \end{equation}
    pushing $Y_i < k_{i-}$ more than half the time and contradicting the definition of $k_{i-}$ as the first integer with $P\left(Y_i \le k_{i-}\right) \ge 0.5$.

    \Needspace{4\baselineskip}
    \textbf{Case 2:}
    \begin{equation}
        \widehat{y} = g\left(\eta\right) \ge k_{i-} + 1.
    \end{equation}
    Then with probability exceeding $0.5$,
    \begin{equation}
        \xi_i \ge g^{-1}\left(k_{i-} + 1\right),
    \end{equation}
    forcing $Y_i$ to exceed $k_{i-}$ more than half the time. Again, this contradicts the minimality of $k_{i-}$.

    Since neither case can occur, we conclude
    \begin{equation}
        k_{i-} \le \widehat{y} < k_{i-} + 1.
    \end{equation}
    Hence, the pseudo-median always lies between the two adjacent integers that bracket the true median, differing by strictly less than one count. Consequently, $\widehat{y}$ is a reliable surrogate for the median and is significantly easier to evaluate or interpret.
    
        
        
        
        
        


        
    
\bibliographystyle{unsrt}


    \newpage
    \clearpage
    
    \setcounter{page}{1}
    \setcounter{figure}{0}
    \setcounter{table}{0}
    \markboth{}{}
    \renewcommand{\restoreapp}{}
    \renewcommand\appendixname{Web Appendix}
    \renewcommand{\figurename}{WEB FIGURE}
    \renewcommand{\tablename}{WEB TABLE}
    \renewcommand{\headrulewidth}{0pt}
    \renewcommand{\thesection}{\Alph{section}}
    \renewcommand{\thesubsection}{\thesection.\arabic{subsection}}
    
    \newpage
    \clearpage
    
    \begin{appendices}
        
        {\Large\bf Addressing outliers in mixed-effects logistic regression: a more robust modeling approach}
        
        \vskip 1.0cm
        
        {\normalsize Divan A. Burger, Sean van der Merwe, Emmanuel Lesaffre}
        
        \vskip 4.5truecm
        
        \begin{center}
            \noindent
            {\Large\bf Supporting Information}
        \end{center}

        \newpage

        \section{Bayesian specification} \label{sec:BAYESSPEC}

        The prior distributions are specified to ensure a weakly informative setup, allowing the data to inform the posterior distributions predominantly.
    
        For each component of the vector of fixed effects $\bm{\beta}$ of each regression model, we assign a normal prior distribution with mean $0$ and variance $1000$:
        \begin{equation}
            \bm{\beta} \sim \mathcal{N}\left(0, 1000\right).
        \end{equation}
        The parameter $\sigma$ in both the binomial-logit-$t$ and binomial-logit-normal models is assigned a half-$t$ prior with location $0$, scale $4$, and $2$ degrees of freedom. This is specified as
        \begin{equation}
            \sigma \sim \text{Half-}t\left(0, 4, 2\right).
        \end{equation}
        Conversely, for the beta-binomial model, the dispersion parameter $\delta$ is assigned a gamma prior with both shape and rate parameters set to $0.001$:
        \begin{equation}
            \delta \sim \text{Gamma}\left(0.001, 0.001\right).
        \end{equation}
        The prior for the degrees of freedom of the binomial-logit-$t$ model, $\nu$, utilizes a hierarchical structure formed as a mixture of an exponential and gamma distribution to achieve a heavier-tailed prior compared to a simple gamma distribution like $\text{Gamma}\left(0.001, 0.001\right)$. This heavier-tailed prior is designed to capture the uncertainty in $\nu$ better and is closer to the Jeffreys prior for the $t$-distribution's degrees of freedom regarding its tail behavior, making it more uninformative. The hierarchical prior for $\nu$ is specified as follows, based on the methodology by Simpson et al. \citep{SIMPSON2017}:
        \begin{equation}
            \epsilon \sim \text{Exp}\left(0.25\right),
        \end{equation}
        \begin{equation}
            \nu \sim \text{Gamma}\left(2, \epsilon\right),
        \end{equation}
        where $\epsilon$, the rate parameter for the gamma distribution, is itself determined by an exponential distribution. Here, the rate parameter for the exponential distribution is set to 0.25, and the shape parameter is $2$ for the gamma distribution.
    
        We specify the MGH-$t$ prior distribution of Huang and Wand \citep{HUANG2013} for each regression model's variance-covariance matrix $\bm{\Sigma}$ as a more appropriate alternative to the conventional inverse Wishart distribution. The MGH-$t$ prior distribution of $\bm{\Sigma}$ is expressed as a mixture representation of gamma distributions for the diagonal entries of the diagonal matrix $\bm{\Psi} = \text{diag}\left(\psi_1, \ldots, \psi_x, \ldots, \psi_d\right)$, and a Wishart distribution for $\bm{\Sigma}^{-1}$ with inverse scale matrix $4\bm{\Psi}$ and degrees of freedom $d + 1$ \citep{burger2023robust}. The gamma distribution parameters for $\psi_x$ ($x=1,\ldots,d$) are set to 0.5 for the shape and $\sfrac{1}{2500}$ for the rate, configuring the diagonal entries of $\bm{\Psi}$. This configuration results in the specification of the half-$t$ prior distribution with a location parameter of 0, a scale parameter of $2500$, and $2$ degrees of freedom for the standard deviation terms in $\bm{\Sigma}$ and a uniform prior distribution, namely $\text{Uniform}\left(-1, 1\right)$, for the correlation terms.
    
        Markov Chain Monte Carlo (MCMC) methods enable drawing samples from the joint posterior distribution of model parameters, the product of all likelihoods, and prior distributions. Software such as \verb|JAGS| \citep{PLUMMER2003} is typically utilized to facilitate the Gibbs sampling process.
  
        \section{Kullback-Leibler divergence} \label{sec:KL_METHOD}
    
        We assess the influence of individual observations on model fits using leave-one-out cross-validation to assess model adequacy, as described by Wang and Luo \citep{WANG2016}. Specifically, we measure the K-L divergence to evaluate the impact of including versus omitting each data point in the dataset.

        Let $\bm{\alpha}$ denote the collection of parameters. For the binomial model, $\bm{\alpha}$ includes only the fixed effects vector $\bm{\beta}$. For the beta-binomial model, $\bm{\alpha} = \left(\bm{\beta}^\prime, \delta\right)^\prime$, where $\delta$ represents the overdispersion parameter. In the binomial-logit-normal model, $\bm{\alpha} = \left(\bm{\beta}^\prime, \sigma^2\right)^\prime$, where $\sigma^2$ is the scale parameter. Lastly, for the binomial-logit-$t$ model, $\bm{\alpha} = \left(\bm{\beta}^\prime, \sigma^2, \nu\right)^\prime$, where $\nu$ is the degrees of freedom parameter of the $t$-distribution. Let $\bm{\Theta}$ represent the complete set of model parameters, including both fixed and random effects.
    
        Let $\bm{\alpha}^{\left(k\right)}$ and $\bm{u}_{i}^{\left(k\right)}$ respectively represent the $k$\textsuperscript{th} posterior sample for $\bm{\alpha}$ and $\bm{u}_i$ ($k=1,\ldots,K$), and $\bm{y}$ the vector containing $y_{ij}$ for all $i = 1, \ldots, I$ and $j = 1, \ldots, J_i$.
        
        Let $P\left(\bm{\Theta} \left| \bm{y}\right.\right)$ represent the posterior distribution of $\bm{\Theta}$ for all $\bm{y}$ (complete dataset), and $P\left(\bm{\Theta} \left| \bm{y}_{\left[ij\right]}\right.\right)$ the posterior distribution of $\bm{\Theta}$ with observation $y_{ij}$ excluded. The Monte Carlo estimate of the K-L divergence between $P\left(\bm{\Theta} \left|\bm{y}\right.\right)$ and $P\left(\bm{\Theta} \left|\bm{y}_{\left[ij\right]}\right.\right)$ under regression model $R$ is given by:
        \begin{flalign}
            \text{KL}_R \left(P\left(\bm{\Theta} \left|\bm{y}\right.\right), P\left(\bm{\Theta} \left|\bm{y}_{\left[ij\right]}\right.\right)\right) & = \log \left\{ \frac{1}{K} \sum_{k=1}^{K} \left[P\left(y_{ij} \left|\bm{\alpha}^{\left(k\right)}, \bm{u}_i^{\left(k\right)} \right.\right)\right]^{-1} \right\} \\
            & + \frac{1}{K} \sum_{k=1}^{K} \log \left[ P\left(y_{ij} \left|\bm{\alpha}^{\left(k\right)}, \bm{u}_i^{\left(k\right)} \right.\right) \right].
        \end{flalign}
        We calculate the K-L divergence estimates for each observation $i$ and $j$ to determine whether $y_{ij}$ under regression model $R$ is influential, effectively identifying data points that significantly affect the parameter estimates. Following the approach of Tomazella et al. \citep{TOMAZELLA2021}, we consider $y_{ij}$ influential if
        \begin{equation}
            0.5 \left(1 + \sqrt{1 - \exp \left[ -2\text{KL}_R \left(P\left(\bm{\Theta} \left|\bm{y}\right.\right), P\left(\bm{\Theta} \left|\bm{y}_{\left[ij\right]}\right.\right)\right) \right]}\right) \geq 0.99.
        \end{equation}

        \section{Residual analysis} \label{sec:MODEL_RESID}
        
        We utilized a simulation-based residual analysis approach to assess our models' adequacy. This method generates scaled residuals by comparing the observed data with the posterior predictive distribution, conditioning on the random effects associated with each observation. The scaled residuals, which range between 0 and 1, provide a probabilistic measure of model fit.
        
        We further diagnosed model fit through several checks: the Kolmogorov-Smirnov test for uniformity of the residual distribution, a parametric bootstrap test to detect potential overdispersion or underdispersion, and a binomial test to identify any observations significantly deviating from the model's expectations. The residuals were also visually inspected using QQ plots. These checks were performed using the \verb|R| package \verb|DHARMa| \citep{hartig2022package}.

        \clearpage
        
        \section{Regression fits} \label{sec:ADDIT_OUTPUT}

        \begin{figure}[htbp]
            \centering
            \includegraphics[trim=0.0cm 0.0cm 0.0cm 0.0cm, clip=true, scale=0.8]{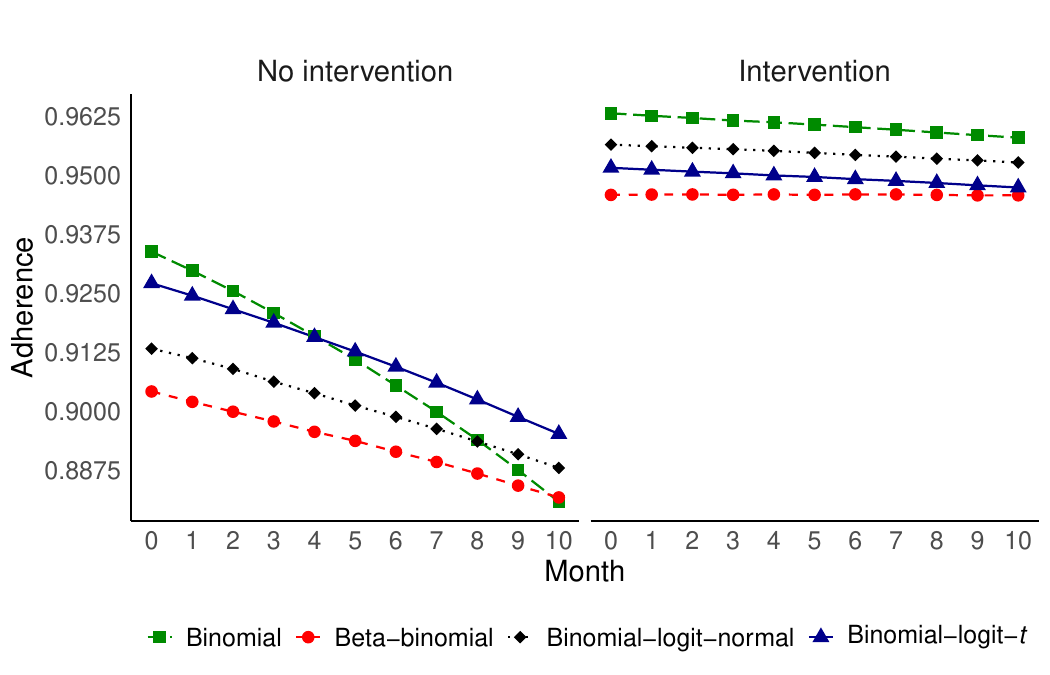}
            \caption{Adherence proportion estimates under the binomial, beta-binomial, binomial-logit-normal, and binomial-logit-$t$ models. The left panel shows patients in the non-intervention group, and the right panel shows those in the intervention group. Green squares with long-dashed lines represent the binomial model, red circles with dashed lines represent the beta-binomial model, black diamonds with dotted lines represent the binomial-logit-normal model, and blue triangles with solid lines represent the binomial-logit-$t$ model. The adherence proportion estimates from the four models are similar across both groups and all time points.}
            \label{fig:All_models_mean_profiles}
        \end{figure}

        \begin{landscape}

        \setlength{\tabcolsep}{0.12cm}

        \begin{table}[ht]
        \centering
        \caption{Parameter estimates and highest posterior density intervals for medication adherence.} \label{tab:posterior_summary}
        {\footnotesize
        \begin{tabular}{lS[table-format=3.2]S[table-format=3.2]S[table-format=3.2]lS[table-format=3.2]S[table-format=3.2]S[table-format=3.2]lS[table-format=3.2]S[table-format=3.2]S[table-format=3.2]lS[table-format=3.2]S[table-format=3.2]S[table-format=3.2]}
        \hline
        & \multicolumn{3}{c}{{\bf Binomial}} & & \multicolumn{3}{c}{{\bf Beta-binomial}} & & \multicolumn{3}{c}{{\bf Binomial-logit-normal}} & & \multicolumn{3}{c}{{\bf Binomial-logit-$\bm t$}} \\
        \cline{2-4} \cline{6-8} \cline{10-12} \cline{14-16}
        {\bf Parameter} & {\bf PE} & \multicolumn{2}{c}{{\bf 95\% HPD}} & & {\bf PE} & \multicolumn{2}{c}{{\bf 95\% HPD}} & & {\bf PE} & \multicolumn{2}{c}{{\bf 95\% HPD}} & & {\bf PE} & \multicolumn{2}{c}{{\bf 95\% HPD}} \\
        \hline
        $\beta_0$ & 1.80 & [0.73; & 2.82] &  & 1.37 & [0.57; & 2.26] &  & 1.77 & [0.71; & 2.90] &  & 2.01 & [0.92; & 2.95] \\ 
          $\beta_{\text{group}}$ & 0.62 & [0.11; & 1.13] &  & 0.61 & [0.20; & 1.08] &  & 0.81 & [0.25; & 1.31] &  & 0.55 & [0.13; & 0.98] \\ 
          $\beta_{\text{time}}$ & -0.07 & [-0.11; & -0.02] &  & -0.02 & [-0.06; & 0.01] &  & -0.03 & [-0.07; & 0.01] &  & -0.05 & [-0.08; & -0.01] \\ 
          $\beta_{\text{tx}}$ & 0.05 & [-0.01; & 0.11] &  & 0.02 & [-0.03; & 0.08] &  & 0.02 & [-0.04; & 0.08] &  & 0.04 & [-0.01; & 0.09] \\ 
          $\beta_{\text{age}}$ & 0.01 & [-0.00; & 0.03] &  & 0.01 & [0.00; & 0.03] &  & 0.02 & [-0.00; & 0.03] &  & 0.02 & [0.00; & 0.03] \\ 
          $\tau_{a_0}$ & 0.40 & [0.33; & 0.47] &  & 0.59 & [0.46; & 0.72] &  & 0.40 & [0.33; & 0.50] &  & 0.51 & [0.40; & 0.64] \\ 
          $\tau_{a_0, a_1}$ & 2.50 & [1.98; & 3.10] &  & 4.21 & [2.97; & 5.67] &  & 2.99 & [2.16; & 4.11] &  & 2.52 & [1.43; & 3.77] \\ 
          $\tau_{a_1}$ & 28.03 & [22.65; & 33.61] &  & 67.79 & [49.37; & 91.71] &  & 47.05 & [33.68; & 63.78] &  & 56.35 & [37.97; & 76.81] \\ 
          $\delta$ &  &  &  &  & 10.01 & [9.05; & 11.15] &  &  &  &  &  &  &  &  \\ 
          $\sigma$ &  &  &  &  &  &  &  &  & 1.05 & [0.99; & 1.10] &  & 0.43 & [0.36; & 0.51] \\ 
          $\nu$ &  &  &  &  &  &  &  &  &  &  &  &  & 1.57 & [1.30; & 1.86] \\ 
          WAIC & 14950.92 &  &  &  & 12201.39 &  &  &  & 11761.90 &  &  &  & 11665.86 &  &  \\ 
        \hline
        \end{tabular}
        }
        \begin{flushleft}
            {\it Definitions/Abbreviations}: $\bm{\Sigma}^{-1} = \begin{bmatrix} \tau_{a_0} & \tau_{a_0, a_1} \\ \tau_{a_0, a_1} & \tau_{a_1} \end{bmatrix}$, where $\tau_{a_0}$ and $\tau_{a_1}$ are the inverse variances for the intercept and slope random effects, and $\tau_{a_0, a_1}$ is the inverse covariance between them; PE = Posterior estimate; HPD = Highest posterior density; WAIC = Watanabe-Akaike information criterion
        \end{flushleft}
        \end{table}
            
        \end{landscape}

        \begin{figure}[!t]
            \centering
            \subfloat[Binomial]{\includegraphics[trim = 0cm 0cm 1.1cm 2.1cm, clip = true, scale = 0.65] {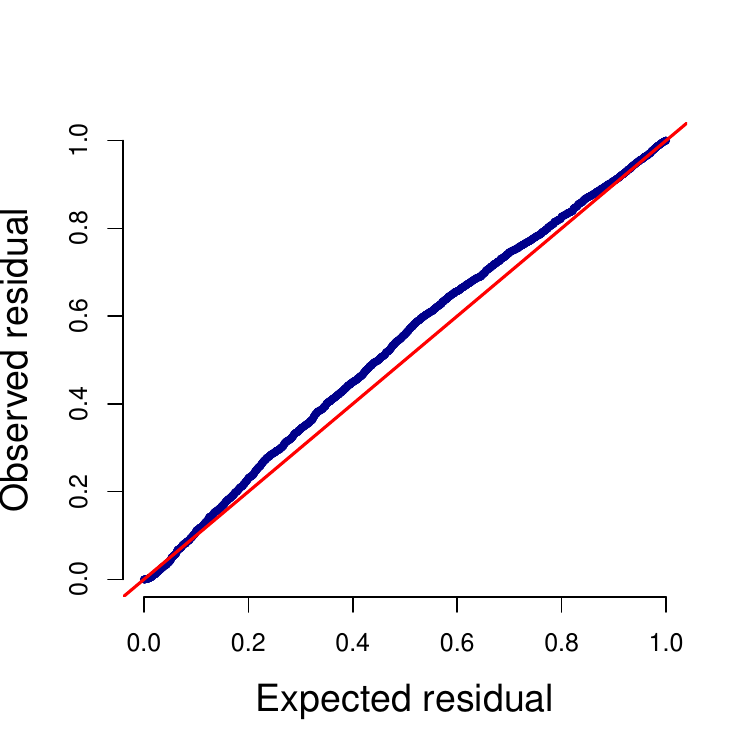}}
            \subfloat[Beta-binomial]{\includegraphics[trim = 0cm 0cm 1.1cm 2.1cm, clip = true, scale = 0.65] {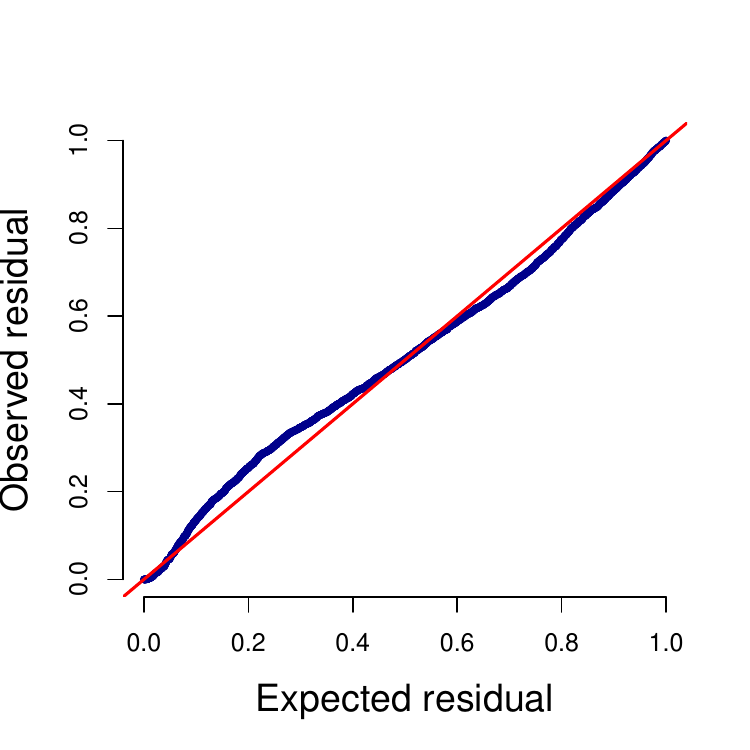}} \\
            \subfloat[Binomial-logit-normal]{\includegraphics[trim = 0cm 0cm 1.1cm 2.1cm, clip = true, scale = 0.65] {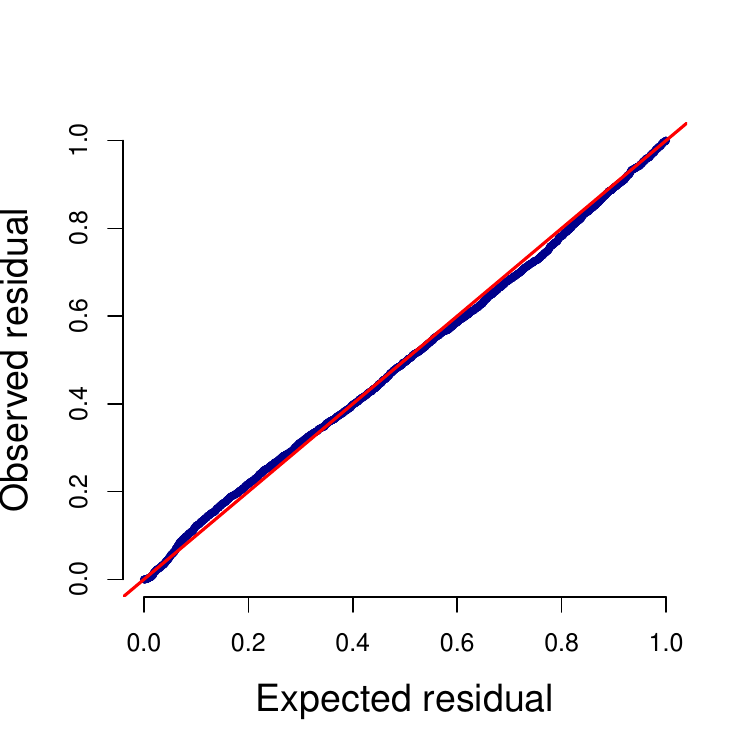}}
            \subfloat[Binomial-logit-$t$]{\includegraphics[trim = 0cm 0cm 1.1cm 2.1cm, clip = true, scale = 0.65] {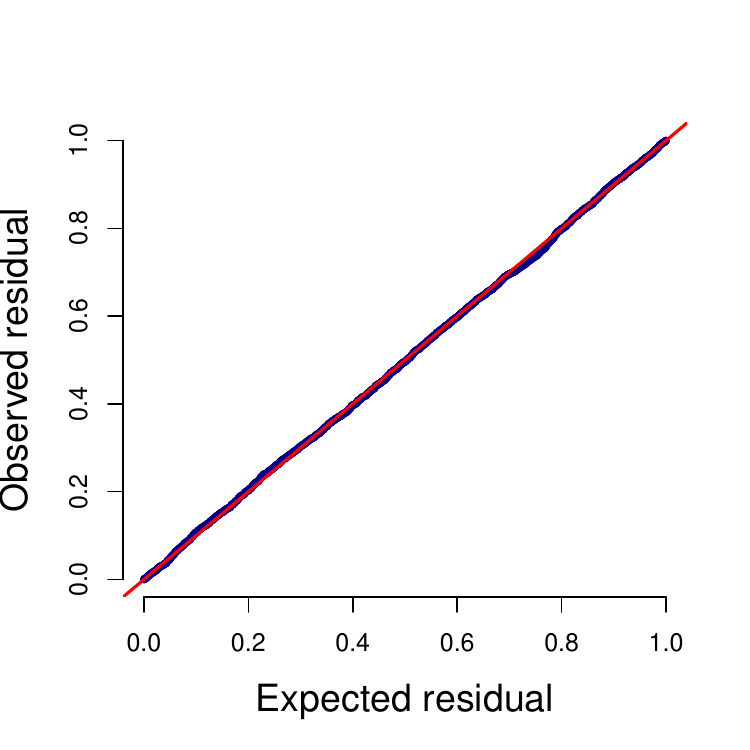}} \\
            \caption{Residual diagnostics across four different models: binomial-logit-$t$, binomial-logit-normal, beta-binomial, and binomial. The QQ plots display the uniformity of the scaled residuals; a closer alignment to the 45-degree red line indicates a better fit. The binomial-logit-$t$ model shows the best fit, with residuals closely aligned across the expected range. The binomial model exhibits the most significant deviations, indicating potential issues with the model's fit and overdispersion. The beta-binomial and binomial-logit-normal models show intermediate performance, with slight deviations.}
            \label{fig:Residual_check}
        \end{figure}

        \section{Simulation study details} \label{sec:APP_SIM_STUDY}

        \subsection{Model performance simulation study} \label{sec:MODEL_PERFORM}

        In this section, we present the simulation study designed to evaluate the performance of the binomial-logit-$t$ model for bounded counts, as detailed in \autoref{sec:MIXED_MODELS}, under different scenarios of overdispersion and outliers.

        The simulation study parameters are chosen to mimic the medication adherence dataset over a 300-day period, with $m = 30$ for the binomial-logit-$t$ model.
    
        The linear predictor $\eta_{ij}$ used in the simulation study is specified as
        \begin{equation}
            \eta_{ij} = \beta_0 + a_{0i} + a_{1i} \cdot \text{time}_{ij} + \beta_{\text{time}} \cdot \text{time}_{ij} + \beta_{\text{tx}} \cdot \text{group}_i \cdot \text{time}_{ij} + \beta_{\text{cnt}} \cdot \text{cov}_i,
        \end{equation}
        where $a_{0i}$ and $a_{1i}$ are the random intercept and slope for cluster $i$, and $\text{group}_i$ and $\text{cov}_i$ represent the treatment group and a continuous covariate, respectively, corresponding to cluster $i$. The random effects vector $\bm{u}_i = \begin{pmatrix} a_{0i} \\ a_{1i} \end{pmatrix}$ follows a bivariate normal distribution with mean vector $\bm{0}$ and covariance matrix $\bm{\Sigma}$, denoted as $\bm{u}_i \sim \mathcal{N}_2\left(\bm{0}, \bm{\Sigma}\right)$. The covariance matrix $\bm{\Sigma}$ is given by
        \begin{equation}
            \bm{\Sigma} = \begin{pmatrix}
            \sigma_{a_0}^2 & \rho_{a_0, a_1} \sigma_{a_0} \sigma_{a_1} \\
            \rho_{a_0, a_1} \sigma_{a_0} \sigma_{a_1} & \sigma_{a_1}^2
            \end{pmatrix},
        \end{equation}
        where $\sigma_{a_0}$ and $\sigma_{a_1}$ are the standard deviations of the random intercept and slope, respectively, and $\rho_{a_0, a_1}$ is the correlation between them.

        We consider four scenarios to evaluate model performance under different conditions of overdispersion and outliers: (i)~$\sigma = 5$ \& $\nu = 5$, (ii)~$\sigma = 5$ \& $\nu = 30$, (iii)~$\sigma = 2$ \& $\nu = 5$, and (iv)~$\sigma = 2$ \& $\nu = 30$. Each of these scenarios is evaluated for two different values of the treatment effect parameter: $\beta_{\text{tx}} = 0.1$ and $\beta_{\text{tx}} = 0.3$. Additionally, for each combination of scenario and $\beta_{\text{tx}}$ value, we consider two different numbers of clusters: 100 clusters ($I = 100$) and 300 clusters ($I = 300$). For all scenarios, the remaining parameters are set as $\sigma_{a_0} = 0.4$, $\sigma_{a_1} = 0.3$, $\rho_{a_0, a_1} = 0.5$, $\beta_0 = 1.5$, and $\beta_{\text{time}} = -0.05$. {\color{red} Web} \autoref{tab:sim_scenario} summarizes all 16 parameter combinations.

        We considered 11 observations ($J_i = 11$) such that $j = 1, \ldots, 11$ and $\text{time}_{ij} = j - 1$. The continuous covariate ($\text{cov}_i$) was simulated from a normal distribution with a mean of 60 and a standard deviation of 5 ($\text{cov}_i \sim \mathcal{N}\left(60, 5\right)$). The treatment group ($\text{group}_i$) was assigned randomly to each subject, simulating the random assignment of subjects to either an intervention or control group ($\text{group}_i \sim \text{Bernoulli}\left(0.5\right)$).

        The evaluation metrics for the binomial-logit-$t$ model under these scenarios include the bias of parameter estimates and the 95\% HPD interval coverage. We simulated 1 000 datasets for each of the 16 scenarios described.

        The simulation study results, including bias and HPD interval coverage, are summarized in {\color{red} Web} \autoref{tab:sim_results}. The results indicate that the bias is small across all simulations investigated. Additionally, the 95\% HPD interval coverage is close to the nominal level.

        \newpage

        \tabcolsep 0.2cm
        \begin{table}[t!]
            \centering
            \caption{Simulation scenarios.} \label{tab:sim_scenario}
            \begin{tabular}{lcccccccccc}
            \toprule
            \textbf{Simulation ID} & $\bm{I}$ & $\bm{\beta_0}$ & $\bm{\beta_{\text{time}}}$ & $\bm{\beta_{\text{tx}}}$ & $\bm{\beta_{\text{cnt}}}$ & $\bm{\omega}$ & $\bm{\nu}$ & $\bm{\sigma_{a_0}}$ & $\bm{\sigma_{a_1}}$ & $\bm{\rho_{a_{01}}}$ \\

            \midrule
            1 & 100 & 1.5 & -0.05 & 0.1 & 0.02 & 0.2 & 5 & 0.4 & 0.3 & -0.5\\
            2 & 100 & 1.5 & -0.05 & 0.1 & 0.02 & 0.5 & 5 & 0.4 & 0.3 & -0.5\\
            3 & 100 & 1.5 & -0.05 & 0.1 & 0.02 & 0.2 & 30 & 0.4 & 0.3 & -0.5\\
            4 & 100 & 1.5 & -0.05 & 0.1 & 0.02 & 0.5 & 30 & 0.4 & 0.3 & -0.5\\
            5 & 100 & 1.5 & -0.05 & 0.3 & 0.02 & 0.2 & 5 & 0.4 & 0.3 & -0.5\\
            6 & 100 & 1.5 & -0.05 & 0.3 & 0.02 & 0.5 & 5 & 0.4 & 0.3 & -0.5\\
            7 & 100 & 1.5 & -0.05 & 0.3 & 0.02 & 0.2 & 30 & 0.4 & 0.3 & -0.5\\
            8 & 100 & 1.5 & -0.05 & 0.3 & 0.02 & 0.5 & 30 & 0.4 & 0.3 & -0.5\\
            9 & 300 & 1.5 & -0.05 & 0.1 & 0.02 & 0.2 & 5 & 0.4 & 0.3 & -0.5\\
            10 & 300 & 1.5 & -0.05 & 0.1 & 0.02 & 0.5 & 5 & 0.4 & 0.3 & -0.5\\
            11 & 300 & 1.5 & -0.05 & 0.1 & 0.02 & 0.2 & 30 & 0.4 & 0.3 & -0.5\\
            12 & 300 & 1.5 & -0.05 & 0.1 & 0.02 & 0.5 & 30 & 0.4 & 0.3 & -0.5\\
            13 & 300 & 1.5 & -0.05 & 0.3 & 0.02 & 0.2 & 5 & 0.4 & 0.3 & -0.5\\
            14 & 300 & 1.5 & -0.05 & 0.3 & 0.02 & 0.5 & 5 & 0.4 & 0.3 & -0.5\\
            15 & 300 & 1.5 & -0.05 & 0.3 & 0.02 & 0.2 & 30 & 0.4 & 0.3 & -0.5\\
            16 & 300 & 1.5 & -0.05 & 0.3 & 0.02 & 0.5 & 30 & 0.4 & 0.3 & -0.5\\
            \bottomrule
            \end{tabular}
        \end{table}
        \begin{flushleft}
            {\it Definitions/Abbreviations}: $\omega = 1/\sigma$
        \end{flushleft}

        \newpage
        
        \begin{longtable}{llS[table-format=3.4]S[table-format=3.4]S[table-format=3.3]}
          \caption{Model performance simulation results}
          \label{tab:sim_results}\\
          \toprule
          {\bf Simulation ID} & {\bf Parameter} & {\bf True value} & {\bf Bias} & {\bf 95\% CP} \\
          \midrule
          \endfirsthead
        
          \caption[]{\textit{(continued from previous page)}} \\
          \toprule
          {\bf Simulation ID} & {\bf Parameter} & {\bf True value} & {\bf Bias} & {\bf 95\% CP} \\
          \midrule
          \endhead
        
          \midrule
          \multicolumn{5}{r}{\textit{Continued on next page}}\\
          \midrule
          \endfoot
        
          \bottomrule
          \endlastfoot
        
        1 & $\beta_0$ & 1.50 & 0.0008 & 0.944\\
         & $\beta_{\text{time}}$ & -0.05 & -0.0017 & 0.953\\
         & $\beta_{\text{tx}}$ & 0.10 & 0.0013 & 0.937\\
         & $\beta_{\text{cnt}}$ & 0.02 & 0.0000 & 0.949\\
        2 & $\beta_0$ & 1.50 & -0.0164 & 0.927\\*
         & $\beta_{\text{time}}$ & -0.05 & 0.0010 & 0.951\\
         & $\beta_{\text{tx}}$ & 0.10 & -0.0020 & 0.960\\
         & $\beta_{\text{cnt}}$ & 0.02 & 0.0004 & 0.933\\
        3 & $\beta_0$ & 1.50 & -0.0175 & 0.958\\
         & $\beta_{\text{time}}$ & -0.05 & 0.0015 & 0.946\\
         & $\beta_{\text{tx}}$ & 0.10 & -0.0023 & 0.953\\
         & $\beta_{\text{cnt}}$ & 0.02 & 0.0003 & 0.962\\
        4 & $\beta_0$ & 1.50 & -0.0344 & 0.949\\*
         & $\beta_{\text{time}}$ & -0.05 & 0.0005 & 0.941\\*
         & $\beta_{\text{tx}}$ & 0.10 & -0.0019 & 0.946\\*
         & $\beta_{\text{cnt}}$ & 0.02 & 0.0006 & 0.944\\
        5 & $\beta_0$ & 1.50 & 0.0118 & 0.941\\
         & $\beta_{\text{time}}$ & -0.05 & 0.0009 & 0.945\\
         & $\beta_{\text{tx}}$ & 0.30 & -0.0016 & 0.946\\
         & $\beta_{\text{cnt}}$ & 0.02 & -0.0001 & 0.941\\
        6 & $\beta_0$ & 1.50 & -0.0074 & 0.950\\
         & $\beta_{\text{time}}$ & -0.05 & -0.0004 & 0.943\\
         & $\beta_{\text{tx}}$ & 0.30 & 0.0003 & 0.945\\
         & $\beta_{\text{cnt}}$ & 0.02 & 0.0002 & 0.947\\
        7 & $\beta_0$ & 1.50 & -0.0047 & 0.955\\
         & $\beta_{\text{time}}$ & -0.05 & 0.0001 & 0.959\\
         & $\beta_{\text{tx}}$ & 0.30 & 0.0015 & 0.947\\
         & $\beta_{\text{cnt}}$ & 0.02 & 0.0001 & 0.955\\
        8 & $\beta_0$ & 1.50 & 0.0243 & 0.946\\*
         & $\beta_{\text{time}}$ & -0.05 & -0.0017 & 0.951\\*
         & $\beta_{\text{tx}}$ & 0.30 & 0.0019 & 0.954\\*
         & $\beta_{\text{cnt}}$ & 0.02 & -0.0004 & 0.948\\
        9 & $\beta_0$ & 1.50 & -0.0079 & 0.951\\
         & $\beta_{\text{time}}$ & -0.05 & 0.0004 & 0.957\\
         & $\beta_{\text{tx}}$ & 0.10 & -0.0009 & 0.950\\
         & $\beta_{\text{cnt}}$ & 0.02 & 0.0001 & 0.952\\
        10 & $\beta_0$ & 1.50 & -0.0020 & 0.941\\*
         & $\beta_{\text{time}}$ & -0.05 & -0.0010 & 0.957\\
         & $\beta_{\text{tx}}$ & 0.10 & 0.0007 & 0.952\\
         & $\beta_{\text{cnt}}$ & 0.02 & 0.0001 & 0.943\\
        11 & $\beta_0$ & 1.50 & 0.0243 & 0.937\\
         & $\beta_{\text{time}}$ & -0.05 & -0.0006 & 0.958\\
         & $\beta_{\text{tx}}$ & 0.10 & 0.0017 & 0.941\\
         & $\beta_{\text{cnt}}$ & 0.02 & -0.0004 & 0.936\\
        12 & $\beta_0$ & 1.50 & 0.0104 & 0.940\\*
         & $\beta_{\text{time}}$ & -0.05 & -0.0006 & 0.952\\
         & $\beta_{\text{tx}}$ & 0.10 & 0.0007 & 0.944\\
         & $\beta_{\text{cnt}}$ & 0.02 & -0.0002 & 0.943\\
        13 & $\beta_0$ & 1.50 & 0.0050 & 0.945\\
         & $\beta_{\text{time}}$ & -0.05 & -0.0007 & 0.943\\
         & $\beta_{\text{tx}}$ & 0.30 & 0.0000 & 0.950\\
         & $\beta_{\text{cnt}}$ & 0.02 & 0.0000 & 0.944\\
        14 & $\beta_0$ & 1.50 & 0.0046 & 0.947\\
         & $\beta_{\text{time}}$ & -0.05 & 0.0012 & 0.940\\
         & $\beta_{\text{tx}}$ & 0.30 & -0.0004 & 0.949\\
         & $\beta_{\text{cnt}}$ & 0.02 & -0.0001 & 0.945\\
        15 & $\beta_0$ & 1.50 & -0.0070 & 0.952\\*
         & $\beta_{\text{time}}$ & -0.05 & 0.0005 & 0.937\\*
         & $\beta_{\text{tx}}$ & 0.30 & 0.0026 & 0.956\\*
         & $\beta_{\text{cnt}}$ & 0.02 & 0.0001 & 0.947\\
        16 & $\beta_0$ & 1.50 & -0.0112 & 0.950\\*
         & $\beta_{\text{time}}$ & -0.05 & -0.0004 & 0.947\\*
         & $\beta_{\text{tx}}$ & 0.30 & -0.0002 & 0.954\\*
         & $\beta_{\text{cnt}}$ & 0.02 & 0.0003 & 0.942\\*
        \end{longtable}
        \begin{flushleft}
            {\footnotesize {\it Notes}: Bias is the mean of the differences between the posterior estimates and the true parameter values. 95\% CP (coverage probability) is the proportion of times the 95\% HPD intervals contain the true parameter values.}
        \end{flushleft}

        \subsection{Contamination simulation study}

        In this section, we build on the model performance simulation study to assess the robustness of the binomial-logit-normal and binomial-logit-$t$ models under varying levels of data contamination. Following the general scheme described in \autoref{sec:MODEL_PERFORM}, we made key modifications to introduce contamination by generating data under the binomial-logit-normal model and then introducing heavy tails, expecting the binomial-logit-$t$ model to demonstrate greater robustness to outliers compared to the binomial-logit-normal model.

        Contamination was introduced by first generating the response variable $y_{ij}$ using the binomial-logit-normal model and then randomly selecting a proportion of these responses based on the contamination rate; selected responses were replaced with extreme values sampled from either the lower tail (values 0, 1, 2) or the upper tail (values $m-2$, $m-1$, $m$) of the binomial-logit-normal distribution, with the same rate applied to both tails, resulting in a mixture of the binomial-logit-normal distribution and two uniform distributions.

        The contamination simulation study used the following parameters: contamination rates $\in \left\{0, 0.02, 0.10\right\}$, $\sigma = 5$, $\sigma_{a_0} = 0.4$, $\sigma_{a_1} = 0.3$, $\rho_{a_0, a_1} = -0.5$, $\beta_0 = 1.5$, $\beta_{\text{time}} = -0.05$, $\beta_{\text{tx}} = 0.1$, $\beta_{\text{cnt}} = 0.02$, $I = 50$, and $m = 30$. We used the same data structure as in the model performance simulation study, with $J_i = 11$ observations per cluster, where $j = 1, \ldots, 11$ and $\text{time}_{ij} = j - 1$, and the covariates $\text{cov}_i$ and $\text{group}_i$ were simulated using the same approach.

        The simulation characteristics assessed include bias, RMSE, average 95\% HPD interval length, and their coverage.

        The simulation results presented in {\color{red} Web} \autoref{tab:sim_contamination} demonstrate that the binomial-logit-$t$ model outperforms the binomial-logit-normal model in the presence of data contamination, particularly in terms of RMSE and the length of the 95\% HPD interval. The binomial-logit-$t$ model exhibited smaller RMSE and shorter HPD intervals, indicating more precise estimates, especially in scenarios with higher contamination rates. Bias and coverage were similar between the two models across all scenarios, consistent with the fact that the scenarios resulted in a symmetrical distribution, minimizing the potential for serious bias issues under contamination. Notably, the shorter HPD intervals with the binomial-logit-$t$ model suggest a gain in statistical power. On the other hand, the binomial-logit-normal model showed greater sensitivity to outliers, with larger RMSE and wider HPD intervals under higher contamination levels. These results suggest that while both models perform comparably in terms of bias and coverage, the binomial-logit-$t$ model provides greater robustness, precision, and statistical power in contaminated data scenarios.

        \newpage

        \begin{landscape}
            \begin{longtable}{lllS[table-format=2.4]S[table-format=2.4]S[table-format=2.4]S[table-format=2.4]S[table-format=1.3]}
            \caption{Data contamination simulation results.} \label{tab:sim_contamination} \\
            \toprule
            {\bf Rate} & {\bf Model} & {\bf Parameter} & {\bf True value} & {\bf Bias} & {\bf RMSE} & {\bf HPD len.} & {\bf 95\% CP} \\
            \midrule
            \endfirsthead
            
            \caption[]{\textit{(continued from previous page)}} \\
            \toprule
            {\bf Rate} & {\bf Model} & {\bf Parameter} & {\bf True value} & {\bf Bias} & {\bf RMSE} & {\bf HPD len.} & {\bf 95\% CP} \\
            \midrule
            \endhead
            
            \midrule
            \multicolumn{8}{r}{{Continued on next page}} \\
            \midrule
            \endfoot
            
            \bottomrule
            \endlastfoot
            0\% & Binomial-logit-normal & $\beta_0$ & 1.50 & -0.0102 & 0.7378 & 3.6752 & 0.947\\
            & & $\beta_{\text{time}}$ & -0.05 & 0.0013 & 0.0489 & 0.2414 & 0.950\\
            & & $\beta_{\text{tx}}$ & 0.10 & 0.0001 & 0.0622 & 0.3234 & 0.957\\
            & & $\beta_{\text{cnt}}$ & 0.02 & 0.0004 & 0.0123 & 0.0612 & 0.950\\
            & Binomial-logit-$t$ & $\beta_0$ & 1.50 & -0.0129 & 0.7475 & 3.6635 & 0.941\\
            & & $\beta_{\text{time}}$ & -0.05 & 0.0011 & 0.0486 & 0.2410 & 0.952\\
            & & $\beta_{\text{tx}}$ & 0.10 & -0.0001 & 0.0622 & 0.3229 & 0.959\\
            & & $\beta_{\text{cnt}}$ & 0.02 & 0.0004 & 0.0125 & 0.0610 & 0.944\\
            2\% & Binomial-logit-normal & $\beta_0$ & 1.50 & 0.1099 & 0.9106 & 4.6293 & 0.950\\
            & & $\beta_{\text{time}}$ & -0.05 & -0.0088 & 0.0484 & 0.2515 & 0.949\\
            & & $\beta_{\text{tx}}$ & 0.10 & 0.0014 & 0.0643 & 0.3356 & 0.951\\
            & & $\beta_{\text{cnt}}$ & 0.02 & -0.0002 & 0.0152 & 0.0771 & 0.954\\
            & Binomial-logit-$t$ & $\beta_0$ & 1.50 & 0.0212 & 0.7710 & 3.6549 & 0.936\\
            & & $\beta_{\text{time}}$ & -0.05 & -0.0019 & 0.0465 & 0.2408 & 0.951\\
            & & $\beta_{\text{tx}}$ & 0.10 & 0.0008 & 0.0632 & 0.3214 & 0.946\\
            & & $\beta_{\text{cnt}}$ & 0.02 & -0.0002 & 0.0129 & 0.0609 & 0.930\\
            10\% & Binomial-logit-normal & $\beta_0$ & 1.50 & 0.0624 & 1.2825 & 6.8407 & 0.965\\*
            & & $\beta_{\text{time}}$ & -0.05 & -0.0164 & 0.0556 & 0.2607 & 0.936\\*
            & & $\beta_{\text{tx}}$ & 0.10 & -0.0044 & 0.0681 & 0.3428 & 0.946\\*
            & & $\beta_{\text{cnt}}$ & 0.02 & 0.0007 & 0.0213 & 0.1136 & 0.964\\*
            & Binomial-logit-$t$ & $\beta_0$ & 1.50 & 0.0236 & 0.7596 & 3.8115 & 0.944\\*
            & & $\beta_{\text{time}}$ & -0.05 & -0.0072 & 0.0487 & 0.2353 & 0.938\\*
            & & $\beta_{\text{tx}}$ & 0.10 & -0.0034 & 0.0643 & 0.3161 & 0.948\\*
            & & $\beta_{\text{cnt}}$ & 0.02 & 0.0000 & 0.0127 & 0.0635 & 0.943\\
    
        \end{longtable}
        \begin{flushleft}
            {\footnotesize {\it Notes}: Rate refers to the contamination rate applied to the dataset. Bias is the mean of the differences between the posterior estimates and the true parameter values. RMSE (root mean square error) is the square root of the average squared differences between the posterior estimates and the true parameter values. HPD length is the average length of the 95\% highest posterior density (HPD) intervals. 95\% CP (coverage probability) is the proportion of times the 95\% HPD intervals contain the true parameter values.}
        \end{flushleft}
    \end{landscape}
        
    \end{appendices}

\end{document}